\documentclass[JSSM]{imsart}
\arxiv{2101.06237}
\firstpage{1}
\lastpage{1}

\usepackage{array,epsfig,fancyhdr,rotating}
\usepackage{siunitx}
\usepackage[nolists,tablesfirst]{endfloat}%Moving figures at the end of the paper
\usepackage{amsmath}
\usepackage{amssymb, amsfonts}
\usepackage{longtable}
\usepackage{natbib}
\usepackage{bm}
\usepackage{graphicx}
\usepackage{color}
\usepackage{floatrow}
\usepackage{units}
\usepackage{url}
\usepackage{verbatim}
\usepackage{epstopdf}
\usepackage{booktabs,caption,fixltx2e}
\usepackage[colorlinks,citecolor=blue,urlcolor=blue,filecolor=blue,backref=page]{hyperref}%For BA format

\usepackage{placeins,afterpage} % gives FloatBarrier to fix position of Figures
\usepackage{xr}%For cross referencing
\newcommand\plotpath[1] {./figs/#1}

\newcommand{\Var}{\mbox{Var}}

\newcommand{\iid}{\stackrel{\mathrm{iid}}{\sim}}

\newcommand{\bth}{\boldsymbol\theta}
\newcommand{\bka}{\hbox{$\boldsymbol{\kappa$}}}

\newcommand{\bbe}{\hbox{$\boldsymbol{\beta$}}}
\newcommand{\bx} {\mathbf{x}}

\newcommand{\bxy} {\mathbf{u}}
\newcommand{\bxp} {\mathbf{v}}
\newcommand{\ie} {\hbox{\textit{i.e.}}}
\newcommand{\eg} {\hbox{\textit{e.g.}}}
\newcommand{\ech}{\color{black}\rm  }    % end change
\newcommand{\hy}{g}   %It was old \hy    But replaced because Terrance did not like this notation
\newcommand{\hmy}{t}    % It was old \hmy

\newcommand{\is}[1] {#1^{(s)}}%In sample indicator
\newcommand{\sampled}[1] {{#1}^{(s)}}      %In sampled vector indicator

\newcommand{\LS}{LS2019}

\newcommand{\REypop}{{\eta}^{y}}
\newcommand{\REpipop}{{\eta}^{\pi}}

\newcommand{\REyDG}{{\eta}^{y,DG}}

\newcommand{\REyAna}{{\eta}^{y,Ana}}

\newcommand{\Sai}{\hbox{$\mathcal {S}1$}}
\newcommand{\Sc} {\hbox{$\mathcal {S}2$}}
\newcommand{\Sd} {\hbox{$\mathcal {S}3$}}

\renewcommand{\labelenumi}  {\roman{enumi} }

\startlocaldefs%For BA format
\numberwithin{equation}{section}%For BA format
\newtheorem{thm}{Theorem}[section]%For BA format
\endlocaldefs%For BA format

\begin{document}
\begin{frontmatter}
\title{Fully Bayesian Estimation under Dependent and Informative Cluster Sampling}
\runtitle{Bayes Estimation/Informative Cluster Sampling}
%\thankstext{T1}{Footnote to the title with the ``thankstext'' command.}

\begin{aug}
\author{\fnms{Luis G.} \snm{Le\'on-Novelo}\thanksref{addr1}\ead[label=e1]
{Luis.G.LeonNovelo@uth.tmc.edu}}
\and
\author{\fnms{Terrance D.} \snm{Savitsky}\thanksref{addr2}%
\ead[label=e2]{Savitsky.Terrance@bls.gov}%
%\ead[label=u1,url]{http://www.foo.com}
}

\runauthor{Le\'on-Novelo \& Savitsky}

\address[addr1]{
Assistant Professor.
University of Texas Health Science Center at Houston-School of Public Health, 1200 Pressler St. Suite E805, Houston, TX, 77030, USA
    \printead{e1} % print email address of "e1"
  %  \printead*{e2}
}

\address[addr2]{
Senior Research Mathematical Statistician.  
Office of Survey Methods Research,   U.S. Bureau of Labor Statistics,
 Washington, DC, 20212-0001, USA
    \printead{e2}
    %\printead{u1}
}

%\thankstext{t1}{Some comment}
%\thankstext{t2}{First supporter of the project}
%\thankstext{t3}{Second supporter of the project}

\end{aug}

%\maketitle
%Incorporation of Primary Sampling Units and Sampling weigh information into Fully Bayesian Analysis.
\begin{abstract}
Survey data are often collected under multistage sampling designs where units are binned to clusters that are sampled in a first stage.  The unit-indexed population variables of interest are typically dependent within cluster.  We propose a Fully Bayesian method that constructs an exact likelihood for the observed sample to incorporate unit-level marginal sampling weights for performing unbiased inference for population parameters while simultaneously accounting for the dependence induced by sampling clusters of units to produce correct uncertainty quantification.  Our approach parameterizes cluster-indexed random effects in both a marginal model for the response and a conditional model for published, unit-level sampling weights. We compare our method to plug-in Bayesian and frequentist alternatives in a simulation study and demonstrate that our method most closely achieves correct uncertainty quantification for model parameters, including the generating variances for cluster-indexed random effects.  We demonstrate our method in an application with NHANES data.
\\
\\
\textbf{Statement of Significance}\\
We propose a fully Bayesian framework for  parameter estimation of a population model from  survey data obtained via a multi-stage sampling design. Inference incorporates sampling weights.   Our framework delivers estimates that  achieve asymptotically correct uncertainty quantification unlike popular Bayesian and frequentist alternatives.   In particular, our method provides asymptotically unbiased point and variance estimates under the sampling of clusters of units. This type of sampling design is common in national and large surveys.

\end{abstract}

\end{frontmatter}

%\def\thefigure{\arabic{figure}}
%\def\thetable{\arabic{table}}

%\renewcommand{\theequation}{\thesection.\arabic{equation}}

%\fontsize{12}{14pt plus.8pt minus .6pt}\selectfont

%\setcounter{section}{0} %***
%\setcounter{equation}{0} %-1

\section{Introduction}
Inference with data from a complex sampling scheme, such as that collected by the National Health and Nutrition Examination Survey (NHANES), requires consideration of the sampling design. A common multistage sampling scheme in public survey datasets is formulated as:
\begin{enumerate}
\item Divide survey population into $H$ strata. 
\item Each stratum is assigned $N_h$ clusters of individuals called
\textit{primary stage sampling units} (PSUs) from which $J_h$ PSUs are selected. 
PSU $hj$ is selected with probability
$\pi_{1hj}$. By design, at least one PSU is selected in each stratum, \ie\ $J_h\geq 1, \forall h$.

\item Within each selected PSU, $n_{hj}$ individuals are sampled out of the total $N_{hj}$ population units nested in the PSU. 
Each individual or last stage unit is sampled with probability $\pi_{i\mid hj}$, $i=1,\dots,N_{hj}$. 
\end{enumerate}
The indices $i,j,h$ index individual, PSU and stratum, respectively.
The marginal probability of including an individual in the sample is then 
$\pi_{ihj}^\prime=\pi_{i\mid hj}\pi_{1hj}$. 

In addition to sampling clusters of dependent individuals, both clusters and individuals-within-clusters are typically selected with unequal sampling inclusion probabilities in order to improve estimation power for a population subgroup or to reduce variance of a global estimator. The sample inclusion probabilities are constructed to be correlated with or ``informative" about the response variable of interest to reduce variance of the estimator.  On the one hand, stratification reduces the standard error (SE) of estimates while, on the other hand, clustering tends to increase the standard error since clustering induces dependence and is used for convenience and to reduce cost. Utilizing unequal inclusion probabilities can reduce the variance of the estimator where a subset of units is highly influential for the estimator of interest, such as is the case where larger-sized employers drive the estimation of total employment for the Current Employment Statistics survey administered by the U.S. Bureau of Labor Statistics; more often, the use of unequal inclusion probabilities tends to increase the variance of the estimator due to the variation in the information about the population reflected in observed samples.  Ignoring PSU and unequal sampling inclusion probabilities underestimates the SE because of the dependence among individuals within a PSU and the variation of information about the population reflected in informative samples drawn from it.

The statistical analyst receives variables of interest for each survey participant
along with the stratum and PSU identifiers
to which s/he belongs, as well as sampling weights, $w_{ihj}\propto 1/\pi_{ihj}$. The inclusion probability, $\pi_{ihj}$,    
is proportional to $\pi_{ihj}^\prime$ after adjusting for oversampling of subpopulations and nonresponse.

In the context of NHANES, a stratum is defined by the intersection of geography with concentrations of minority populations and a
PSU is constructed as a county or a group of geographically continuous counties. Secondary and tertiary stage sampling units include segments (contiguous census blocks) and households. The final unit is an eligible participant in the selected household. NHANES released masked stratum and PSU information to protect participant's privacy. 
Every 2-year NHANES-data cycle \citep{CDCNHANESSurveyDesign} releases information 
 obtained from $H=15$ strata with $J_h=2$ PSUs per stratum.

In this paper, we focus on a two-stage sampling design that excludes strata for both our simulation study and application, in the sequel, without loss of generality since the inclusion of strata would be expected to improve estimation by reducing variability across realized samples.  Our two-stage sampling design of focus is characterized by the first stage sampling of PSUs and, subsequent, second stage sampling of units.  We employ a fully Bayesian estimation approach that co-models the response variable of interest and the marginal inclusion probabilities as introduced in \cite{leon2019fully}, hereafter referred to \LS.  We extend their approach by constructing PSU-indexed random effects specified in both the marginal model for the response variable and the conditional model for the sampling inclusion probabilities. 
 
Since  our sampling design does not utilize strata, we do not consider subindex $h$ that indexes strata in the discussion above.
% NHANES, inference requires consideration of both: 
%1) the first two stages of the sampling design -  strata and primary sampling units (PSU) -  to which the survey participant belongs, 
%and 2) the sampling weights for each observation that adjust for oversampling of subpopulations and nonresponse.
%NHANES provides (masked) strata and PSU memberships of each participant.
%PSUs (counties or clusters of them) are selected from strata defined by geography and concentrations of minority populations. Most strata contain two PSUs.
%Measurements of the individuals in the same strata/PSU are presumed to be associated. 
%Information of participants in 30 PSU  are released every 2-year NHANES-data cycle (\cite{CDCNHANESSurveyDesign}). 
Sampled individual $ij$ denotes individual $i \in \{1,\ldots,n_{j}\}$ in cluster $j \in \{1,\ldots,J_{pop}\}$ included in the sample, where $J_{pop}$ denotes the total number of PSUs in the population.  
Let $J \leq J_{pop}$ denote the number of PSUs actually sampled. 
We assume that the sampling weight, $w_{ij}$, is proportional to  
the inverse marginal inclusion probability, $\pi_{ij}$, of individual $ij$ being
included in the sample; or $\pi_{ij}\propto 1/w_{ij}$.  We denote the vector of predictors associated to individual $ij$ as $\bx_{ij}$.  

The data analyst aims to estimate the parameters, $\bth$, of a \emph{population} model, $p(y\mid\bth,\bx)$,
that they specify from these data. 
Relabeling PSU indices in the sample so they run from $1,\ldots,J$,
the analyst observes sample of size, $n=\sum_{j=1}^J n_j$, and the associated variables, 
$\{\is{y}_{ij},\sampled{\bx}_{ij},\is{\pi}_{ij}\propto 1/\is{w}_{ij},j\}_{i=1,\dots,n_j,j=1,\dots,J}$ 
with $n_j$ the number of participants from PSU $j$ and superindex $(s)$ denotes in the sample.
By contrast, $y_{ij}$ without superindex $(s)$ denotes a 
response of an individual in the survey population but not, necessarily, a survey participant included in the observed sample.

The probability of inclusion of each PSU (denoted as $\pi_{1j}$ in point ii, above) is unknown to the analyst because it is not typically published for the observed sample, though the PSU inclusion probabilities are used to construct published unit marginal inclusion probabilities, %$\pi_i=Pr(ST_i) Pr(PSU_i\mid ST_i)\times \dots$, and, 
 (such that inclusion probabilities within the same PSU tend to be more similar or correlated), but the  dependence of units in the same PSU may not be fully accounted for by the dependence on their inclusion probabilities.

A sampling design is informative for inference about individuals within a group when $y_{ij}\not\perp \pi_{ij}\mid \bx_{ij}$.  
A sampling design will also be informative for PSUs in the case that 
$\bar{y}_{\cdot j} - \bar{y} = (1/N_{j})\mathop{\sum}_{i=1}^{N_{j}} y_{ij} - \bar{y} \not\perp \pi_{1j}\mid \bar{\bx}_{j}$ with
 $\bar{y}$ the population mean response and  
$\bar{\bx}_{j}=(1/N_j) \sum_{i=1}^{N_j} \bx_{ij} $. Even if a sampling design is not informative for individuals and/or groups, however, there are typically scale effects induced by within group dependence that must be accounted for to produce correct uncertainty quantification.    

\begin{comment}
\LS\ propose a  model-based Bayesian approach appropriate under informative sampling 
that incorporates the sampling weights into the model 
by modelling both the response given the parameter of interest and the inclusion probability given the response, $\pi_{ij}\mid y_{ij}$. The main advantages of this approach is that it yields 
(1) consistent point estimates [LS2019] (point estimates converge in probability to true values),
(2)  credible intervals that achieve nominal (frequentist) coverage, and
(3) robust inference against mis-specification of $\pi_{ij}\mid y_{ij}$. 
\end{comment}

\LS\ only %\cite{leon2019fully} 
focus on fixed effect models and ignore
the dependence induced by the sampling design; that is, both
association among the responses within the same PSU (that we label, dep-$y$), and  
possible association among inclusion probabilities within the same PSU (that we label, dep-$\pi$).  
This paper extends the approach of  \LS\ %\cite{leon2019fully} 
to account for these associations via mixed effect models. More specifically,
we include PSU-specific random effects (PSU-REs)
in both the model for the
responses and in the model for the inclusion probabilities.

\cite{makela2018bayesian} propose, as we do, Bayesian inference 
under a
two-stage sampling design.
\begin{comment}
in particular, they consider 
the case where  clusters/PSUs are selected
with probability, \ie $\pi_{1j}$, proportional to a measure of PSU size 
(that is commonly the number of individuals in the PSU). They require $\pi_{1j}$  to be available and published to the data analyst for the sampled PSUs.  They assume that individuals nested in PSUs are drawn under simple random sampling (SRS) in a second stage.
\end{comment}
Their estimation focus is on the population mean or proportion.
%Their estimate is based on the outcome prediction of the 
%(1) non sampled individuals in the selected PSUs, 
%(2) the mean (or proportion) in the non selected PSUs,
%(3) and their measure of size.  
%Under our approach estimation of the population mean would require us 
%to assume the model 
%know the distribution of the vector of predictors, ${\bx}_{ij}$, in the population.
By contrast, we focus on estimation of model parameters and assume that 
the analyst does not know $\pi_{1j}$ (because it is not published), but instead only knows $\pi_{ij}$ (up to a multiplying constant) for the sampled individuals, as is the case for NHANES data.  We do not assume that individuals within the sampled PSUs are selected under SRS, but allow for informativeness.

We introduce our general approach in Section \ref{sec:inclusionofPSU}, though in the rest of the paper  
we focus on the linear regression setting for ease-of-illustration. Competing methods are summarized in this section as well.  
In Section \ref{sec:simulationstudy},
we show via simulation that our approach yields
credible intervals with nominal (frequentist) coverage, 
while the competing methods do not in some simulation scenarios.
In Section \ref{sec:applications} we demonstrate our approach by applying it to an 
NHANES dataset 
%The first application models the relationship between %percentage of body fat to body mass index while the second 
%The application  estimates 
 to estimate the daily kilocalorie consumption of persons in different demographic groups in the U.S. population.
Inference under our Fully Bayes approach is compared against inference under competing plug-in Bayesian and frequentist methods. 
We provide a final discussion section and an Appendix 
containing details not discussed, but referred to in the main paper.

\section{Review of \LS}\label{sec:review}
\LS\ introduces the inclusion probabilities into the Bayesian paradigm by assuming them to be random.
 In this section we review their approach before we extend it to include PSU information, in the next section.
The superpopulation approach in \LS\ assumes that the finite population of size $N$, 
$(y_1,\pi_1,\bx_1),\dots (y_N,\pi_N,\bx_N)$ is a realization of   
\begin{equation}\label{eq:population}
(y_i,\pi_i)\mid \bx_i,\bth,\bka \sim p(y_i,\pi_i\mid \bx_i,\bth,\bka)=  
p(\pi_i\mid y_i,\bx_i,\bka)\ p(y_i\mid \bx_i,\bth), \quad i=1,\dots,N.
\end{equation}
Here, $y_i$ is the response for individual $i$ with vector of covariates $\bx_i$ and 
$\pi_i\in [0,1]$ is a proper survey sampling inclusion  probability for individual $i$ being sampled. 
It is assumed that
$(y_i,\pi_i)\perp (y_{i^\prime},\pi_{i^\prime})\mid \bx_i,\bx_{i^\prime},\bth,\bka$, 
for $i\neq i^\prime$, and $\bx_i$  is assumed known; that is, the unit responses and inclusion probabilities are conditionally (on the model parameters) independent.
Note also that \eqref{eq:population} above presumes that 
$\pi_i\perp \bth \mid y_i,\bx_i,\bka$ and
$y_i\perp \bka \mid \bx_i,\bth$; that is, the parameters for the models for the response and weights are \emph{a priori} independent. 
The population parameter $\bth$ determines the relationship between $\bx_i$ and $y_i$, and is of main interest. The parameter $\bka$ is a nuisance parameter that allows modeling the association between $\pi_i$ and $y_i$, though we later see in our simulation study section that it provides insight on the informativeness of the sampling design for a particular response variable of interest.
%Define $\delta_i$ as the indicator of the individual $i$ in the informative sample, 
%\ie, $\delta_i=1$ if $i$ is a participant and $0$ otherwise.
The informative sample of size $n$ is drawn so that 
$P[\hbox{individual $i$ in sample}]=\pi_i$, a proper sampling inclusion probability. 
Bayes theorem implies,
\begin{align}\label{eq:samplingdist}
  p(y_{i},\pi_{i}\vert \bx_{i}, \bth, \bka,&\hbox{individual $i$ in sample})\\ 
  =&\frac
{\mbox{Pr}(\hbox{individual $i$ in sample} \vert y_{i},\pi_{i},\bx_{i}, \bth, \bka  )\times p(y_i,\pi_i\vert \bx_i, \bth, \bka)   \nonumber}
{\hbox{denominator}}
%\\
%f(y_i, \pi_i | x_i, I_i = 1) \times Pr(I_i = 1 | x_i)  =& Pr(I_i = 1 | y_i, x_i, \pi_i) \times p(\pi_i|y_i,x_i) \\
 %     &\times p(y_i|x_i)
\end{align}
By the way the informative sample is drawn, and the  population model in \eqref{eq:population}, the numerator in
\eqref{eq:samplingdist}
is 
\begin{equation}\label{eq:numerator}
\pi_i \times
p(\pi_i\mid y_i,\bx_i,\bka)\ p(y_i\mid \bx_i,\bth)
\end{equation}
The denominator is obtained by integrating out $(y_i,\pi_i)$ in the numerator, 
\begin{equation}\label{eq:denominator}
\int \pi_i^\star
p(\pi_i^\star \mid y_i^\star,\bx_i,\bka)\ p(y_i^\star\mid \bx_i,\bth)\, d\pi_i^\star dy_i^\star=
E_{y_i^\star\mid \bx_i,\bth}\left[E\left(\pi_i^\star\mid y_i^\star,\bx_i,\bka\right)\right]
\end{equation}
The superindex $\star$ is used to distinguish the quantities integrated out from the ones in the numerator. 
Plugging \eqref{eq:numerator} and \eqref{eq:denominator} in \eqref{eq:samplingdist} we obtain Equation (5) in \LS, 
and also Equation (7.1) in
\cite{pfeffermann1998parametric},  
given by,
\begin{equation}\label{eq:IScorrection}
p_s(y_{i},\pi_{i}\vert \bx_{i}, \bth, \bka)=
\left\{\frac
{\pi_i\,   p(\pi_i\vert y_i,\bx_i,\bka)   }
{E_{y_i^\star\vert \bx_i,\bth}\left[E(\pi_i^\star\vert y_i^\star, \bx_i,  \bka) \right]}\right\}
\times p( y_i\vert \bx_i, \bth)
\end{equation}
where the LHS, $p_s(\cdots\mid \cdots)$, denotes the joint distribution of $(y_i,\pi_i)$ 
conditioned on the individual $i$ being in the sample, \ie, the LHS of \eqref{eq:samplingdist} is equal to
$p(\dots\mid \cdots,\hbox{individual $i$ in sample})$. Inference is based on this \emph{exact} likelihood for the observed sample with,
\begin{equation*}\label{eq:likelihood}
\ell(\bth,\bka;\sampled{y},\sampled{\pi},\sampled{\bx})=\prod_{i=1}^n\left[p_s(\is{y_i},\is{\pi}_i\mid \sampled{x_i},\bth,\bka) \right]
\end{equation*}
where the superindex $(s)$ is used to emphasize that these are the values observed in the sample. We also  relabel the index $i$ running from $1,\dots,N$ in the population so it runs from $1,\dots,n$ in the sample. 
A Bayesian inference model is completed by 
assigning priors to $\bth$ and $\bka$.

Note that under noninformative sampling, \ie\ when $y_i\perp \pi_i\mid \bx_i$, 
the quantity between curvy brackets in \eqref{eq:IScorrection} does not depend on $y_i$ and therefore inference on $\bth$ does not depend on the inclusion probabilities, or the $\pi_i$s. In other words, inference using \eqref{eq:IScorrection} is the same as if treating the sample as an SRS from the model $y_i\sim p(y_i\mid \bx_i,\bth)$.
For the informative sampling case,
in theory, we can assume any distribution for $y_i\mid \bx_i,\bth$ and $\pi_i\mid y_i,\bx_i,\bka$.
In practice, the calculation of $E_{y_i^\star\vert \bx_i,\bth}[\cdots]$ in \eqref{eq:IScorrection}
is a computational bottleneck. Theorem 1 in \LS, stated below,  provides 
conditions to obtain a closed form for this expected value. 

Let $\bxp_i$ and $\bxy_i$ be subvectors of $\bx_i$, the covariates used to specify the conditional distribution of $\pi_i\mid y_i,\bx_i,\bka$
and $y_i\mid \bx_i,\bth$, respectively; that is,
$\pi_i\mid y_i,\bx_i,\bka \sim \pi_i \mid y_i,\bxp_i,\bka$ and
 $y_i\mid \bx_i,\bth\sim y_i\mid \bxy_i,\bth$.  Note that we allow for $\bxp_i$ and $\bxy_i$ to have common covariates.
Let $\hbox{normal}(x\mid\mu,s^2)$ denote the normal distribution pdf with mean $\mu$ and variance $s^2$ evaluated at $x$, and $\hbox{lognormal}(\cdot\mid\mu,s^2)$ denote the lognormal pdf, so that
$X\sim     \hbox{lognormal}(\mu,s^2)$ is equivalent to  $\log X\sim\hbox{normal}(\mu,s^2)$.

\begin{thm}\label{th:closeform} (Theorem 1 in \LS)
If
$p(\pi_i\mid y_i,\bxp_i,\bka) =\emph{lognormal}(\pi_i\mid h(y_i,\bxp_i,\bka),\sigma_{\pi}^2)$,
with the function $h(y_i,\bxp_i,\bka)$ of the form $h(y_i,\bxp_i,\bka)=\hy(y_i,\bxp_i,\bka)+\hmy(\bxp_i,\bka)$ where
%$\bka=(\bka,\bka)$ and
$\sigma_{\pi}^2=\sigma_{\pi}^2(\bka,\bxp_i)$, possibly a function of $(\bka,\bxp_i)$
then
$$
p_s(y_i,\pi_i\mid \bxy_i,\bxp_i,\bth,\bka)=
\frac{\emph{normal}\left(\log \pi_i\mid \hy(y_i,\bxp_i,\bka)+\hmy(\bxp_i,\bka),\sigma_\pi^2\right)}
       {\exp\left\{\hmy(\bxp_i,\bka)+\sigma^2_\pi/2\right\}  \times M_y(\bka;\bxy_i,\bxp_i,\bth)    }
\times p(y_i\mid \bxy_i,\bth)\nonumber
$$
 with $M_y(\bka;\bxy_i,\bxp_i,\bth):=E_{y^\star_i\mid \bxy_i,\bth}\left[\exp\left\{\hy(y^\star_i,\bxp_i,\bka)\right\}\right]$.
\end{thm}
If both $M_y$ and $p(y_i\mid\cdots)$ admit closed form expressions, then $p_s(y_i,\pi_i\mid\cdots)$ has a closed form, as well;
for example, when %$\bka$ has dimension 1 and
$\hy(y_i,\bxp_i,\bka)=\kappa_y y_i$ 
where  
$\kappa_y$ an element of the parameter vector, $\bka$, with
$\kappa_y \in \mathbb{R}$, 
then $M_y(\bka;\bxy_i,\bxp_i,\bth)$ is the moment generating function (MGF) of $y_i\mid \bth$ evaluated at $\kappa_y$, which may have a closed form defined on $\mathbb{R}$. This implies a closed form for
$p_s(y_i,\pi_i\mid\cdots)$.  
Analogously, we may consider an interaction between $y$ and $\bxp$, using 
$\hy(y_i,\bxp_i,\bka)=(\kappa_y+\bxp_i^t \bka_\bxp) y_i\equiv r y_i$ with
$\bka=(\kappa_y,\bka_\bxp,\sigma_\pi^2)$. In this case, we achieve,  $M_y(r;\cdots)$, which is the MGF evaluated at $r$. 
As mentioned in \LS, the assumption of a lognormal distribution for $\pi_{i}$ is mathematically appealing. The inclusion probability, $\propto \pi_i$, for individual, $i$, is composed from the product of inclusion probabilities of selection across the 
stages of the multistage survey design. 
If each of these stagewise probabilities are {lognormal} then their product, $\propto\pi_i$, is {lognormal} as well. This is particularly helpful in the setting that includes PSUs, discussed in next section.

For implementation, we \emph{observe} sampled $\{(\sampled{\pi}_i,\sampled{y}_i)\}_{i=1,\dots,n}$ and we estimate the exact posterior distributions for the population model parameters on the observed sample.
Under our lognormal conditional model for $\pi_i$s, there is no restriction imposed on $\sum_{i=1}^n \sampled{\pi}_i$, such that we may normalize the $\sampled{\pi}_s$ to any positive constant, $\sum_{i=1}^n \sampled{\pi}_i = c$,
as long as $h(y_i,\bxp_i,\bka)=\kappa_0+\dots$ includes an intercept parameter that we label $\kappa_0$. 
Since $\pi_i\sim\hbox{lognormal}(\kappa_0+\dots,\dots)$ is equivalent to
$\pi_i/c\sim\hbox{lognormal}(\kappa_0-\log c+\dots,\dots)$ such that the estimated intercept  is either
$\kappa_0$, or a shifted version, $\kappa_0-\log c$, inference is unaffected.

%The normalizing constant, $c$, factors and cancels in the numerator and denominator as the scale induced by $c$ is captured in estimation of the parameters of the model for $\pi_{i}$; in particular, for $(\bm{\kappa},\sigma_{\pi}^{2})$.
%and under a lognormal prior $\log\pi_{i}$, $\log c$ will be absorbed into the intercept.

\section{Inclusion of PSU Information into Fully Bayesian approach}\label{sec:inclusionofPSU}

In Subsection \ref{subsec:reviewLN}, we extend the approach in \LS, reiviewed in 
Section \ref{sec:review},
that co-models the response variable and sampling weights, modifying their notation by adding cluster-indexed parameters,
in preparation to include  PSU information into the analysis in 
Subsection  \ref{subsec:includingPSU} to capture within PSU dependence in the response and sample inclusion probabilities. In Subsection \ref{subsec:LRM} we introduce the Fully Bayes joint population model for the response and the sample inclusion probabilities
in the linear regression case. In Subsections \ref{subsec:pseudo} and \ref{subsec:freq} we briefly review competing approaches to analyze informative samples that we will compare in a simulation study.
%In the sequel we may call $\pi_{ij}$, $\pi_{1j}$ and $\pi_{i\mid j}$ 
%probabilities of inclusion in the sample when we mean proportional to the

\subsection{Extend Joint Population Model to incorporate PSU-indexed Parameters \label{subsec:reviewLN}}
%We will work under the superpopulation model paradigm.
We assume a population with a total of $J_{pop}$ PSUs and
size $N=\sum_{j=1}^J N_j$ with  $N_j$ the number of population individuals or units in PSU $j$. 
More specifically, the population consists of
$$
\begin{array}{rl}
\underbrace{(y_{1,1},\pi_{1\mid1},\bx_{1,1}),\dots,(y_{N_1,1},\pi_{N_1\mid1},\bx_{N_1,1}  )}_{\hbox{PSU }1},&\dots,
\underbrace{(y_{1,j},\pi_{1\mid j},\bx_{1,j}),\dots,(y_{N_j,j},\pi_{N_j\mid j},\bx_{N_j,j})}_{\hbox{PSU }j},\dots\\
\multicolumn{2}{c}{
\underbrace{(y_{1,J_{pop}},\pi_{1\mid  J_{pop}},\bx_{1,J_{pop}}),
\dots,(y_{N_{J_{pop}},J_{pop}},\pi_{N_J{_{pop}} \mid J_{pop}},\bx_{N_{J_{pop}},J_{pop}}  )}_{\hbox{PSU }J_{pop}} 
} 
\end{array}
$$
 and also of,
$
\pi_{11},\pi_{12},\dots,\pi_{1j},\dots ,\pi_{1J_{pop}}>0
$
with  $\pi_{i\mid j}\in (0,1],~ \forall i,j$.
%Notice that, as before, 
%\bch since  we will include an intercept parameter
%in the model for $\pi_{ij}$, \ech
%we do not impose any restriction on $\sum_{i=1}^{N_j} %\pi_{i|j}$ or
%$\sum_{j=1}^{J_{pop}} \pi_{1j}$.
%We also have the unstandardized PSU inclusion probabilities 
%$$.
 The sample of size $n=\sum_{j=1}^J n_j$ 
 (with $n_j$ and $J$ specified by the survey sampler)
 is drawn in two steps:
  \begin{itemize}
      \item{Step 1: PSU sampling.} 
      Sample $J$ different PSUs $j_1,\dots,j_J\in\{1,\dots,J_{pop}\}$ so that
      \begin{equation*}%\label{eq:PSUsamplingstep}
      Pr[\hbox{PSU } j \hbox{ is in the sample}] = \pi_{1j}
      \end{equation*}

    \item {Step 2: Sampling of individuals.} Within each  PSU in observed sample $j\in\{j_1,\dots,j_J\}$,   
    draw 
    $n_j$ different individuals 
    so that individual $i$ (in the sampled PSU $j$) is in the sample with probability  
    $$
      P[\hbox{Individual } {ij} \hbox{ is in the sample}\mid \hbox{PSU } j\hbox { is in the sample}]= \pi_{i\mid j} %\frac{\pi_{ji}}{\pi_{1j}}
    $$
  \end{itemize}
  and therefore the marginal inclusion probability is proportional to $\pi_{ij}:=\pi_{1j}\pi_{i\mid j}$.

The superpopulation approach assumes 
that the population is a realization of 
the joint distribution for values of the response variable and inclusion probabilities,
\begin{align}\label{eq:jointyandpi}
(y_{ij},\pi_{ij})\mid \bx_{ij},\bth,\REypop_j,\bka,\pi_{1j} \sim&\   
p(y_{ij},\pi_{ij}\vert \bx_{ij}, \bth,\REypop_j,\bka,\pi_{1j})\\%=&p(\pi_i\vert y_i,\bx_i,\bth, \bka) p(y_i\vert \bx_i,\bth, \bka)\\
=&\   p(\pi_{i j}\vert y_{ij},\bx_{ij},\bka,\pi_{1j})\, p(y_{ij}\vert \bx_{ij},\bth,\REypop_j)
\nonumber
\end{align}
We model $\pi_{ij}\mid y_{ij}$ with
$p(\pi_{ij}\vert y_{ij},\bx_{ij},\bka,\pi_{1j})$.  
 The (population) parameter of interest is $\bth$. 
 This construction allows for an informative sampling design by modeling $\pi_{ij}$ conditioned on $y_{ij}$.  While $(\pi_{ij},y_{ij})$ are assumed to be conditionally independent over PSUs $j$ and units $i$, they are unconditionally (on model parameters) dependent under our construction.  We have augmented the parameters used in \LS, given in 
 \eqref{eq:population}, 
 to incorporate 
 $\REypop_j$ and $\pi_{1j}$ that are shared by all observations in PSU $j$. Parameters,
 $\REypop_j$, induce a correlation in the  response for individuals in the same PSU (dep-$y$)
 while $\pi_{1j}$ induces association of marginal inclusion probabilities (dep-$\pi$) among respondents nested in the same PSU.  We will later construct priors on these parameters to define PSU-indexed random effects.
  $\bka$ is a nuisance parameter used to model the inclusion probabilities.
  After relabeling the sampled PSU indices 
  $j_1,\dots j_J$ to $1,\dots J$, and the indices $i$ in the sample to run from $i=1,\dots,n_j$, 
  the sample of size $n=\sum_{j=1}^J n_j$ consists of
   $$%(\sampled{\boldsymbol{y}},\sampled{\boldsymbol{\pi}},\sampled{\boldsymbol{x}})
     \hbox{\emph{data} }:=
   \{\is{y}_{ij},\sampled{\bx}_{ij},\is{\pi}_{ij},j\}_{i=1,\dots,n_j; j=1,\dots,J}$$ 
with 
$j$ indicating from which PSU the individual was sampled,
$n_j$ the number of participants from PSU $j$,  
and
$J$ the total number of  sampled PSUs.
Recall,  superindex $(s)$ denotes in the sample.
The equality in \eqref{eq:jointyandpi} assumes that 
$y_{ij}\perp (\bka,\pi_{1j})\mid \bx_{ij},\bth,\REypop_j$ and 
$\pi_{ij}\perp   (\bth,\REypop_j)  \mid y_{ij},\bx_{ij}, \bka,\pi_{1j}$. 
%Equation \ref{eq:jointyandpi}  implies that, except when $y_{ij} \perp \pi_{ij}\mid \bx_{ij}$,  our sample is informative.

\begin{comment}
Examples of noninformative sample are  
\begin{enumerate}
\item SRS: equivalent to $J_{pop}=J=1$ a and $\pi_{i\mid 1}=1$  for $i=1,\dots,N$.
\item SRS within PSU with PSU sampling probability $\pi_{1j}$ independent of the response, 
equivalent  to $\pi_{1j}\perp y_{ij}\mid \bx_{ij}$ $\forall i$ and $\pi_{i\mid j}=1$.
\end{enumerate}
\end{comment}

%This is,
%response and inclusion probability are not independent, in math,
%$y_{ji} \not\perp \pi_{ji} \mid \bx_{ji},\bth,\bka$.
%Since the marginal probability of including individual $ji$ is $\pi_{ji}$.
We extend \eqref{eq:IScorrection} that captures the joint probability model for the sample by replacing $\bth$ and $\bka$ with
($(\bth,\REypop)$ and $(\bka,\pi_{1j})$) to achieve,
\begin{equation}\label{eq:IScorrectionPSU}
p_s(y_{ij},\pi_{ij}\vert \bx_{ij}, \bth,  \REypop_j,\bka,\pi_{1j})=\frac
{\pi_{ij}\,   p(\pi_{ij}\vert y_{ij},\bx_{ij},\bka,\pi_{1j})   }
{E_{y_{ij}^\star\vert \bx_{ij},\bth,\REypop_j}\left[E(\pi_{ij}^\star\vert y_{ij}^\star, \bx_{ij},  \bka,\pi_{1j}) \right]}
\times p( y_{ij}\vert \bx_{ij}, \bth,\REypop_j)
\end{equation}
The subindex $s$ on  the joint distribution $p_s$ on the LHS denotes that we condition on individual $ij$ being in the sample; that is, 
$p_s(y_{ij},\pi_{ij}\mid \dots)=p(y_{ij},\pi_{ij}\mid \hbox{individual } ij \hbox{ is in the sample},\dots)$. 
In contrast, the distributions on the RHS  are population distributions.
Inference on $\bth$ utilizes the joint likelihood for the observed sample,
 \begin{equation*}%\label{eq:likelihood}
\ell(\bth,\boldsymbol{\REypop},\bka,\boldsymbol{\pi}_1;
%\sampled{\boldsymbol{y}},\sampled{\boldsymbol{\pi}},\sampled{\boldsymbol{x}}
\hbox{\emph{data}}
)=
\prod_{j=1}^J \prod_{i=1}^{n_j} 
\left[p_s(\is{y_{ij}},\is{\pi}_{ij}\mid \sampled{x}_{ij},\bth,\REypop_j,\bka,\pi_{1j}) \right]
\end{equation*}
with $\boldsymbol{\REypop}:=(\REypop_1,\dots,\REypop_J)$, 
$\boldsymbol{\pi}_1:=(\pi_{11},\dots,\pi_{1J})$.
%and the superindex $(s)$ is used to denote quantities in the sample.
Inference for $\bth$ is achieved via the posterior distribution of the model parameters:
\begin{align*}
p_s\left(\bth,\boldsymbol{\REypop},\bka,\boldsymbol{\pi}_1 \mid
%(\sampled{\boldsymbol{y}},\sampled{\boldsymbol{\pi}},\sampled{\boldsymbol{x}})
\hbox{\emph{data}}
\right)\propto& \
\ell\left(\bth,\boldsymbol{\REypop},\bka;
%(\sampled{\boldsymbol{y}},\sampled{\boldsymbol{\pi}},\sampled{\boldsymbol{x}})
\hbox{\emph{data}}
\right)
\times \hbox{Prior}(\bth)           \times \hbox{Prior}(\REypop)
\times \hbox{Prior}(\boldsymbol{\pi}_1)           \times \hbox{Prior}(\bka).
\end{align*}
To obtain a closed form for the likelihood we need a closed form for the expected value
in the denominator in \eqref{eq:IScorrectionPSU}, in turn. 
 Theorem  \ref{th:closeform} (same as Theorem 1 in \LS) is here extended under our extended PSU-indexed parameterization,  ($(\bth,\REypop)$ and $(\bka,\pi_{1j})$), to provide conditions 
that allow a closed form expression of this expected value.
Similar to the set-up for Theorem \ref{th:closeform}, 
let $\bxp$ and $\bxy$ be subvectors of $\bx$, 
the covariates used to specify the conditional distribution of 
$\pi_{ij}\mid y,\bx,\bka,\pi_{1j}$
and $y\mid \bx,\bth,\REypop$, respectively; that is,
$\pi_{ij}\mid y_{ij},\bx_{ij},\bka,\pi_{1j} \sim \pi_{ij} \mid y_{ij},\bxp_{ij},\bka,\pi_{1j}$ and
 $y_{ij}\mid \bx_{ij},\bth,\REypop\sim y_{ij}\mid \bxy_{ij},\bth,\REypop$.  

\begin{thm}\label{th:closeformPSU}
If
$p(\pi_{ij}\mid y_{ij},\bxp_{ij},\bka,\pi_{1j}) =\emph{lognormal}(\pi_{ij}\mid h(y_{ij},\bxp_{ij},\bka,\pi_{1j}),\sigma_{\pi}^2)$,
with the function $h(y_{ij},\bxp_{ij},\bka,\pi_{1j})$ of the form $h(y_{ij},\bxp_{ij},\bka,\pi_{1j})=
              \hy(y_{ij},\bxp_{ij},\bka)+
              \hmy(\bxp_{ij},\bka,\pi_{1j})$ where
%$\bka=(\bka,\bka)$ and
$\sigma_{\pi}^2=\sigma_{\pi}^2(\bxp_{ij},\bka,\pi_{1j})$,
possibly a function of $(\bxp_{ij},\bka,\pi_{1j})$
then
\begin{align*}
p_s(y_{ij},\pi_{ij}\mid \bxy_{ij},\bxp_{ij},\bth,\REypop_j,\bka,\pi_{1j})=&
\frac{\emph{normal}\left(\log \pi_{ij}\mid \hy(y_{ij},\bxp_{ij},\bka)+\hmy(\bxp_{ij},\bka,\pi_{1j}),\sigma_\pi^2\right)}
       {\exp\left\{\hmy(\bxp_{ij},\bka,\pi_{1j})+\sigma^2_\pi/2\right\}  
       \times M_y(\bka;\bxy_{ij},\bxp_{ij},\bth,\REypop_j)    }\\
&\times p(y_{ij}\mid \bxy_{ij},\bth,\REypop_j)
\end{align*}
 with $M_y(\bka;\bxy_{ij},\bxp_{ij},\bth):=
 E_{y_{ij}^\star\mid \bxy_{ij},\bth,\REypop_j}\left[\exp\left\{\hy(y_{ij}^\star,\bxp_{ij},\bka)\right\}\right]$.
\end{thm}
So, analogously to the discussion after Theorem \ref{th:closeform}, 
if $\hy(y,\bxp,\bka)=\kappa_y y$
with $\kappa_y$ depending  on 
$\bka$ and, perhaps, on $\bxp$,
then 
$$M_y(\bka;\bxy_{ij},\bxp_{ij},\bth,\REypop_j):=
E_{y_{ij}^\star\mid \bxy_{ij},\bth,\REypop_j}\left[\exp\left(\kappa_y y_{ij}^\star\right)\right]$$
 is the moment generating function of $y$ evaluated at $\kappa_y$. So when 
both the population distribution of $y$, $p(y\mid \bth,\REypop,\bxy)$, 
has a closed form
and 
the moment generating function has a closed form over the real line, 
then the likelihood, $p_s$, has a closed form, as well. 
%In this manuscript we will work with $\hy(y)=\kappa_y y$ and 
%$\hmy(\bxp_{ji},\bka) := (1,\bxp_{ji}^t)\bka_\bxp)+\REpi_j$, with $\REpi_j$ a PSU specific  

\subsection{Inclusion of PSU Information into Conditional Population Model for Weights \label{subsec:includingPSU}}
The marginal inclusion probability 
of the individual $ij$, $\propto\pi_{ij}$, 
 is the product of the probability of selecting PSU $j$, $\propto\pi_{1j}$, %defined in Equation~\ref{eq:PSUsamplingstep},
 and the probability of selecting the individual $i$ conditioning on PSU being in the sample, $\propto\pi_{i\mid j}$ such that
$\pi_{ij}=\pi_{1j} \pi_{i\mid j}.$
%\quad\hbox{with }\pi_{i\mid j}:=\pi_{ji}/\pi_{1j} 
Therefore,
%The sample $(y_{i},x_{i},\pi_{i},j)$ be the response, set of covariates, inclusion probability and PSU index. 
%We model the inclusion probabilities:
$$
\log \pi_{ij}=\log \pi_{i\mid j}+\log \pi_{1j}
$$
Specifying,   
$\log \pi_{1j}\sim \hbox{normal}(\mu_j,\sigma_{\REpipop}^2)$ where 
$\mu_j$ could depend on PSU covariates 
(e.g. county population, etc) 
but, for simplicity, we assume that it does not and  set $\mu_j=0$.
Choosing a normal distribution for 
$
\log \pi_{i\mid j}\sim \hbox{normal}(
\hy(y,\bxp,\kappa)+\hmy^\prime(\bxp,\bka),\sigma_{\pi}^2)$ yields
\begin{equation}\label{eq:logpiij}
\log \pi_{ij}\mid y_{ij},\bxp_{ij},\bka,\REpipop_j
\sim \hbox{normal}\left(\hy(y_{ij},\bxp_{ij},\bka)+
\hmy^\prime(\bxp_{ij},\bka)+\eta^{\pi}_{j},\sigma_{\pi}^2\right)
\end{equation}
 with $\REpipop_j:=\log \pi_{1j}\iid \hbox{normal}(0,\sigma_{\REpipop}^2)$  PSU-specific random effects. 
 So defining
 $\hmy(\bxp,\bka,\pi_{1j}):=$ $\hmy^\prime(\bxp,\bka)+\log \pi_{1j}=\hmy^\prime(\bxp,\bka)+\REpipop_j$,
 the distribution of $\pi_{ij}$ satisfies the conditions of Theorem \ref{th:closeformPSU}.
 %and 
 %$\bka:=(\bka,\eta^{\pi}_1,\dots,\eta^{\pi}_J,\sigma_{PSU}^2)$ to have the same model as the one in the paper.
 
This set-up is coherent with our assumption that the data analyst does not have information about the PSU-indexed sampling weights for either the population or sampled units because they are not typically published by survey administrators.  Nevertheless, our derivation of \eqref{eq:logpiij} by factoring the marginal inclusion probabilities, $\pi_{ij}$, demonstrates how we may capture within PSU dependence among $(\pi_{ij})$ by inclusion of random effects, $\REpipop_j$.  
%Our Fully Bayes construction has the feature that it does not require availability of PSU-indexed random effects, unlike the pseudolikelihood approach of \citet{williams2021}.
 
 Notice that, as before, 
 since  we  will include an intercept parameter,
$\kappa_0$,
in the model for $\pi_{ij}$ in \eqref{eq:logpiij}, \ie,
$\hmy^\prime(\bxp,\bka)=\kappa_0+\dots$,
we do not impose any restriction on
$\sum_{i=1}^{n_j} \pi_{ij}$ or $\sum_{j=1}^J \sum_{i=1}^{n_j} \pi_{ij}$.
%$ or
%$\sum_{j=1}^{J_{pop}} \pi_{1j}$.
%We also have the unstandardized PSU inclusion probabilities 
%$$.

\subsection{Linear Regression Joint Population Model}\label{subsec:LRM} 
We construct a linear regression model for the population with,
%$p(y_{ij}\mid \bxy_{ij},\bth,\REpipop_j)$, is constructed as,
 \begin{equation}\label{eq:SLR_likelihood}
{y_{ij}\mid \bxy_{ij},\bth,\REypop_j}\sim\text{normal}\left(\bxy_{ij}^t\bbe+\REypop_j ,\sigma_y^2 \right) , 
\quad\hbox{with }\bth=(\bbe,\sigma_y^2)
\end{equation}
with the PSU-specific random effect $\REypop_j$ in \eqref{eq:SLR_likelihood} 
playing the roll of $\REypop_j$ in \eqref{eq:jointyandpi}. 
The conditional population model for inclusion probabilities is specified as in \eqref{eq:logpiij},
with 
\begin{equation}\label{eq:lonnormalpriorforpi}
{\pi_{ij}\mid y_{ij},\bxp_{ij},\bka,\REpipop_j}\sim
\text{lognormal}\Big(\kappa_y y_{ij}+\bxp_{ij}^t \bka_\bxp+\REpipop_j,   \sigma_\pi^2\Big),\quad\hbox{with }                                               \bka=(\kappa_y,\bka_\bxp,\sigma_\pi^2)
\end{equation}
This construction results from setting, $\hy(y_{ij},\bxp_{ij},\bka)=k_y y_{ij}$,
$\hmy(\bxp_{ij},\bka,\pi_{1j})=\bxp_{ij}^t \bka_\bxp+\REpipop_j$ (remember $\REpipop_j=\log \pi_{1j}$), and  
$\sigma_\pi^2(\bka,\bxp_{ij},\pi_{1j})=\sigma_\pi^2$ in ~\eqref{eq:logpiij}.
 Here $\bbe$ and $\bka_\bxp$ are vectors of regression coefficients
 that include an intercept, so the first entry of both $\bxy_{ij}$ and $\bxp_{ij}$ equals 1.
 We select prior distributions,
 %%Multivariate normal
% vague priors are used for $\bbe$, and $\bka$; and we choose
 \begin{equation}\label{eq:priors}
 \begin{array}{c}
 \bbe \sim \hbox{MVN}(\mathbf{0},100 \mathbf{I}), \quad                  
 \bka \sim \hbox{MVN}(\mathbf{0},100 \mathbf{I}), \quad                  
 \REypop_1,\dots,\REypop_J\iid \hbox{normal}(0,\sigma_{\REypop}^2), \\
 \REpipop_1,\dots,\REpipop_J\iid \hbox{normal}(0,\sigma_{\REpipop}^2), \quad \hbox{and} 
 \quad
 \sigma_y,\sigma_\pi,\sigma_{\REypop},\sigma_{\REpipop} \iid \hbox{normal}^+(0,1)
 \end{array}
 \end{equation}
 with $\hbox{normal}^+(m,s^2)$
 denoting a normal distribution with mean $m$ and variance $s^2$ restricted to the positive real line;
 $\hbox{MVN}(\mathbf{m},\bm{\Sigma})$  the multivariate normal distribution with
 mean vector $\mathbf{m}$ and variance-covariance matrix $\bm{\Sigma}$; and
 $\mathbf{I}$  the identity matrix. Since
$y\sim\hbox{normal}(m,s^2)$ admits a closed form expression for 
moment generating function 
$M_y(t)=\exp(tm+t^2 s^2/2)$,
we apply Theorem \ref{th:closeformPSU} to obtain,
\begin{align}\label{eq:LRp_s}
p_s\left(y_{ij},\pi_{ij}\mid \bxy_{ij},\bxp_{ij},\bth,\REypop_j,\bka,\REpipop_j\right)
=&\frac{\hbox{normal}\left(\log \pi_{ij}\mid \kappa_y y_{ij}+\bxp_{ij}^t\bka_\bxp+\REpipop_j,\sigma_\pi^2\right)}
       {\exp\left\{\bxp^t_{ij} \bka_\bxp+\REpipop_j +\sigma^2_\pi/2+  
       \kappa_y (\bxy^t_{ij}\bbe+\REypop_j)+\kappa_y^2\sigma_y^2/2     \right\}}\nonumber \\
&\times \hbox{normal}\left(y_{ij}\mid \bxy^t_{ij}\bbe+\REypop_j ,\sigma^2_y\right)
\end{align}
The implementation of the Gibbs sampler is not straightforward in this case due to non-conjugacy under the exact likelihood in \eqref{eq:LRp_s}. 
To obtain a posterior sample of the model parameters,
we rely on the ``black box'' solver, ``Stan" \citep{carpenter2016stan}, which performs an 
efficiently-mixing Hamiltonian Monte Carlo sampling algorithm with a feature that 
non-conjugate model specifications are readily accommodated. 
%In Stan one can access and add terms to the log of the full conditional by adding the line
%$target+=$.  
%To specify 
%the model in Stan we need to either pass the log of  
%\eqref{eq:LRp_s}, or, equivalently, specif

%The resulting form of the expression for $\log p_s(\dots)$  is provided in the Supplementary Material, Section~
%\ref{subsec:suppleLinReglikelihood}.

\subsection{Pseudolikelihood Approach}\label{subsec:pseudo}
%We discuss other methods to two stage weighted samples.
\cite{10.1214/18-BA1143} propose an approach 
to incorporate sampling weights using a plug-in observed data pseudolikelihood:
\begin{equation}\label{eq:fullpseudo}
\left[\prod_{j=1}^J \prod_{i=1}^{n_j} p(\sampled{y}_{ij}\mid \theta)^ {\sampled{w}_{ij}}\right] \times \prod_{j=1}^J p(\REypop_{j}\mid \sigma^{2}_{\REypop})
\end{equation}
where we start with a pseudolikelihood that exponentiates each observed data likelihood contribution by its marginal unit sampling weight, $\sampled{w}_{ij}$, to re-balance the information in the observed sample to approximate that in the population.  So the observed data pseudolikelihood (in square brackets) is not an exact likelihood, but an approximation for the unobserved population.  We apply the pseudolikelihood approach by augmenting it with the prior for the unobserved, linked random effect to form an augmented pseudolikelihood. Inference on $\bth$ utilizes the pseudoposterior.
In Section \ref{sec:simulationstudy} we label this approach, ``Pseudo".

\cite{10.1214/18-BA1143} standardize marginal individual sampling weights so that\\  $\sum_{j=1}^J\sum_{i=1}^{n_j} \sampled{w}_{ij}=n$ to approximately reflect the amount of posterior uncertainty in the sample. Nevertheless, as in \LS,  
neither do they account for  dep-$y$ nor dep-$\pi$, so resultant credibility intervals are overly optimistic (short).   In a related work, \citet{williams2021} propose to utilize a separate, post-processing step applied to the pseudoposterior samples that produces posterior draws with the sandwich  variance estimator (that depends on $(y_{ij},w_{ij})$) of the pseudo MLE to account for the dependence induced by clustering. The added post processing step to correct the posterior variance to the sandwich form that characterizes the frequentist construction is required because the pseudolikelihood treats the weights as fixed.   In our fully Bayesian approach, 
by contrast, the frequentist sandwich form collapses to the Bayesian estimator for the asymptotic covariance matrix under joint modeling of the response variable and sampling weights under assumption of a joint population generating model \citep{kleijn2012}. 

%$\REypop_j$ is removed from  Equation~\ref{eq:pseudolikelihood} and the pseudposterior reduces to expression in 
%in equation Equation~\ref{eq:pseudoprop} after removing the quantities in 
%curly brackets. 

Related frequentist approaches of \cite{pfeffermann1998weighting}  and 
\citep{rabe2006multilevel} %Rabe-Hesketh and Skrondal (RHS) (2006) 
also employ an augmented likelihood similar to that of 
\eqref{eq:fullpseudo}, but where they also weight the prior for the random effect with the marginal group (PSU)-level weight, $\sampled{w}_{1j}$.  They proceed to integrate out the random effect, 
$\eta^{y}_{j}$, to perform estimation.   
They also focus on consistent estimation of parameters, 
rather than correct uncertainty quantification.

We will see in the sequel that because our approach uses a joint model for $(y_{ij},\pi_{ij}\mid \bx_{ij})$, the asymptotic covariance matrix of the joint posterior, $H^{-1}$, is the same for the MLE such that we achieve correct uncertainty quantification.

In the context of the linear regression in  
\eqref{eq:SLR_likelihood}, the quantity between square brackets in 
\eqref{eq:fullpseudo} matches the likelihood of the weighted regression model
${y_{ij}\mid \bxy_{ij},\bth,\REypop_j}\sim\text{normal}\left(\bxy_{ij}^t\bbe+\REypop_j ,\sigma_y^2/w_{ij}\right)$
(See Appendix \ref{subsec:pseudoandweightedreg} for details.)
So \eqref{eq:fullpseudo} becomes
\begin{equation*}
\left[\prod_{j=1}^J 
\prod_{i=1}^{n_j} \text{normal}\left(\sampled{y}_{ij}\mid \bxy_{ij}^t\bbe+\REypop_j ,\sigma_y^2/\sampled{w}_{ij} \right)
\right] \times \prod_{j=1}^J p(\eta^{y}_{j}\mid \sigma^{2}_{\eta^{y}})
\end{equation*}
This becomes useful under estimation in Stan where one can specify the weighted linear regression model and add the log of $p(\REypop\mid \cdots)$ to the $\log$ of the full conditional for joint sampling of the model parameters.

%If the model for the response has not PSU-specific REs, 
%(we will call this approach pseudo\_WORE in section 

\subsection{Frequentist Approach}\label{subsec:freq}
Frequentist estimation approaches are designed-based, assuming the population is fixed.
The formulation that we highlight employs the pseudolikelihood construction, but without PSU-REs. 
The point estimate of $\bth$, called $\tilde{\bth}_{freq}$, maximizes\\
$p_{pseudo}(\bth; \sampled{y},\sampled{\pi},\sampled{x})=
\prod_{j=1}^J \prod_{i=1}^{n_j}  \left[p(\sampled{y}_{ij}\vert \sampled{\bx}_{ij}, \bth)\right]^{\sampled{w}_{ij}}$
with $\sampled{w}_{ij}\propto 1/\sampled{\pi}_{ij}$ standardized so that 
$\sum_{ij} \sampled{w}_{ij}=n$. 
The PSU indices, together with $p_{pseudo}$, 
are used to
estimate the standard error of 
$\tilde{\bth}_{freq}$ via resampling methods
(\ie, balanced repeated replication, Jack-Knife, Bootstrap) 
or Taylor series linearization. 
The R function svyglm in the R package 
\hbox{survey}~\citep{lumley2019surveyR}
uses the latter (as default) to fit 
common generalized regression models such as 
linear, Poisson, logistic, etc.   

For multiple linear regression with $\bth=(\bbe,\sigma_y^2)$,  
inference, in particular the construction of  confidence regions, 
for the $(p+1)$-dimension vector of regression coefficients $\bbe$ 
(that includes an intercept)  
 is based on the asymptotic result,
$\tilde{\Sigma}^{1/2} (\tilde{\bbe}_{freq}-\bbe)\sim (p+1)\hbox{-variate Student-t}$ with 
degrees of freedom equal to  
$df=\# PSUs-\#Strata$, that represents the design-based degrees of freedom; 
$\tilde{\Sigma}^{1/2}$ is a lower triangular scale matrix such that  $\tilde{\Sigma}^{1/2}\tilde{\Sigma}^{1/2}=\tilde{\Sigma}$ with 
$\tilde{\Sigma}$ the estimate of the variance-covariance matrix of $\tilde{\bbe}_{freq}$.
No stratification is equivalent to 
having one stratum and the degrees of freedom reduces to 
$df=J-1$ (recall, $J:=\#PSUs$). 
This frequentist approach for uncertainty quantification is
similar to the post processing correction of \citet{williams2021} in that the analysis model for the population does not employ a PSU-indexed random effects term; rather, the resampling of clusters captures the dependence within clusters.  Both methods perform nearly identically, in practice, so that we focus on comparing our Fully Bayes approach to this frequentist resampling method in the simulation study that follows.

\section{Simulation
\label{sec:simulationstudy}}
We perform a Monte Carlo simulation study to compare the performance of our fully Bayes method of \eqref{eq:LRp_s} that employs PSU-indexed random effects in both the models for the response and inclusion probabilities to the pseudoposterior and frequentist methods, presented in Sections ~\ref{subsec:pseudo} and ~\ref{subsec:freq}, respectively.  In each Monte Carlo iteration, we generate a population of $J_{pop}$ clusters and $N_{j}$ individuals per cluster. 
%In order to induce dependence in the response variable. 
The response variable is generated proportionally to size-based group and marginal inclusion probabilities to induce informativeness (dependence between the response variable and inclusion probabilities).  We next take a sample of groups and, subsequently, individuals within group.  A clustered simple random sample (cSRS) is also generated from the same population.  The cSRS is included to serve as a gold standard for point estimation and uncertainty quantification (under the population model) and is compared to our model alternatives designed for estimation on the informative sample taken from the same population.  For each population and sample we utilize the Fully Bayes method and associated comparative methods.  We assess the bias, MSE and coverage properties under each model formulation.
 
\subsection{Monte Carlo Simulation Scheme}\label{subsec:MCscheme}
Steps 1-5 describe how the synthetic population dataset is generated, steps 6-7 how the samples are drawn and 8-10 how they are analyzed. We use the superindex `DG' to refer to the data generating (population) model as opposed to the analysis model. In the sequel, gamma$(a,b)$ denotes  the gamma distribution with shape and rate parameters $a$ and  $b$ (\ie, mean $a/b$).
\renewcommand{\labelenumi}{\arabic{enumi}}

\begin{enumerate}
\item %Every PSU has $N_j=10^3$ individuals.
Generate $\pi_{i\mid j}\iid \hbox{gamma}(a_\pi=2,b_\pi=2)$ for $i=1,\dots, (N_j=20)$ individuals nested in PSU, $j=1,\dots,(J_{pop}=10^3)$ total PSUs. The total population size is $J_{pop}\times N_j = 20,000$.

\item Define PSU $j$ inclusion probability $\pi_{1j}^{tem}:=\sum_{i=1}^{N_j}\pi_{i\mid j}$ (therefore $\pi_{1j}^{tem}\iid$  $\hbox{gamma}(N_j a_\pi,b_\pi)$).

\item Standardize $\pi_{1j}:=\pi_{1j}^{tem}/\sum_{j^\prime=1}^{J_{pop}} \pi_{1j^\prime}^{tem}$, so 
$$(\pi_{1,1},\pi_{1,2},\dots,\pi_{1,J_{pop}})\sim\hbox{Dirichlet}\left( a_\pi \times(N_1,N_2,\dots,N_{J_{pop}})\right).$$ 
(Thus, $b_\pi$ does not play a roll on the distribution of $\pi_{1j}$.) 
%Notice that $\pi_{1j}\sim Beta(a_\pi,a_\pi(J-1))$ and, therefore, 
%$$
%var(\pi_{1j})=\frac{a_\pi^2(J-1)}{(J a_\pi)^2 (J a_\pi+1)}=
%\frac{J-1}{J^2 (J a_\pi+1)}\to_{a_\pi\to0} \frac{J-1}{J^2} 
%$$

\item Generate $\REyDG_j\iid \hbox{normal}(0,\sigma_{\REyDG}^2=0.1^2)$  
PSU specific random effects  and  predictor
$x_{ij}\iid \hbox{Uniform}(0,1)$.

\item Generate the response. 
We consider three simulation scenarios to generate the response by different settings for coefficients in the following generating expression,
$$y_{ij}=
\beta_0^{DG}+\beta_1^{DG} x_{ij}+
\beta_{\pi,1} {\pi}_{1j}+
\beta_{\pi,2} {\pi_{i\mid j}}+
\beta_{\REyDG}  \REyDG_j+ 
%\beta_{\pi,\REyDG} (\pi_{1j} \times \REyDG_j)+
\epsilon_{ij}^{DG},$$
with $\epsilon^{DG}_{ij}\iid \hbox{normal}(0,(\sigma_y^{DG})^2)$.  The three scenarios each set the last three regression coefficients, as follows:
\begin{center}
\begingroup
\setlength{\tabcolsep}{4pt} % Default value: 6pt
\begin{tabular}{l| cccc}
Scenario &$\beta_{\pi,1}$&$\beta_{\pi,2}$& $\beta_{\REyDG}$\\%&$\beta_{\pi,\REyDG}$              \\
\hline
{\Sai}: Informative PSU-RE       &$J_{pop}$  & 1     &0\\%  &$J_{pop}$\\
{\Sc}: Non-informative PSU-RE    &0          & 1     &1      \\ 
{\Sd}: No stage is informative   &0          & 0     & 1     \\
\end{tabular}
\endgroup
\end{center}
with $(\sigma_y^{DG})^{2}=0.1^2,\beta_0^{DG}=0,\beta_1^{DG}=1$. Note that, in scenario \Sai,  
$\beta_{\pi,1}=1/E(\pi_{1j})=J_{pop}$ so that $\beta_{\pi,1} E(\pi_{1j})=1$.  Informative random effects are instantiated in Scenario \Sai\ by generating $y_{ij}$ from $\pi_{1j}$, where $\pi_{1j}$, the inclusion probability for PSU, $j$, is equivalent to a PSU-indexed random effect. 
We set the regression coefficient for $\pi_{1j}$ equal to $0$ and that for $\REyDG_j$ equal to $1$ in Scenario \Sc, where we generated random effects as non-informative (uncorrelated with the selection probabilities).

\item Take a clustered simple random sample (cSRS) from the population:
From each population dataset we draw two samples, 
one informative %under the informative sampling design as described in subsection \ref{subsec:reviewLN} 
and the other under two-stage cluster SRS.  Both samples contain $J=30$ PSUs and $n_j=5$  individuals that produces a total sample size of
$n=\sum_{j=1}^J n_j=150$.  Results under cSRS will serve as a  Gold Standard and  will be compared
to the results under comparative methods designed to analyze informative samples.  To implement the clustered random sample we,
\begin{enumerate}
\item Draw an SRS (without replacement) of size $J$ of PSUs indices from $\{1,\dots,J_{pop}\}$.
%, call it  \{j_1,\dots,j_J\}.
\item Within each drawn PSU $j$, obtain a SRS (without replacement) of size $n_j$.
\item Relabel PSU indices to run from $1$ to $J$ and individual indices to
 run from $1$ to $n_j$.
\item The cSRS consists of  $\{(y_{ij},x_{ij},j)\}_{i=1,\dots,n_j;j=1,\dots,J}$
\end{enumerate}

\item
Take an informative sample:
\begin{enumerate}
\item  Draw, without replacement, $J$ PSU indices $j_1,\dots,j_J\in\{1,\dots,J_{pop}\}$ with $Pr(j\in\hbox{sample})=\pi_{1j}$. 
\item For each $j\in\{j_1,\dots,j_J\}$ drawn, sample, without replacement, 
$n_j$ individual indices 
$i\in\{1,\dots, N_{j}\}$  with probability $Pr(i\in \hbox{ sample from PSU }j)=\pi_{i\mid j}/\sum_{i^\prime=1}^{N_j} \pi_{i^\prime\mid j}$.
%Note that the total sample size is $n=\sum_{j=1}^{J} n_j=30\times 5=150$. %Call $\sampled{\pi}_{i\mid j}$ the sampled ${\pi}_{i\mid j}$s.

\item %Obtain the sample $(y_{i\mid j}^{IS},x_{i\mid j}^{IS},\pi_{1j}^{IS},\pi_{i\mid j}^{IS},i,j)$
Define $\pi_{ij}=\pi_{1j}\pi_{i\mid j}$ and
relabel the PSU and individual indices so they run from $1$ to $J$ and from $1$ to $n_j$,
respectively, and add superindex ``$(s)$'' to denote sampled quantities.

\item The informative sample consists of 
$\{(\sampled{y}_{ij},\sampled{x}_{ij},\sampled{\pi}_{ij},j)\}_{ 
i=1,\dots,n_j, j=1,\dots,J}$.
\end{enumerate}

\item 
Analyze the realized informative sample by estimating parameters under the following modeling approaches: 
\begin{enumerate}
    \item \textbf{FULL.both}: Denotes the approach enumerated in Subsection \ref{subsec:includingPSU} that employs PSU-REs in models for both response and inclusion probability; the model for the response includes PSU-indexed random effects with,
    \begin{equation}\label{eq:AnalysisModel}
    y_{ij}=\beta_{0}^{Ana}+\beta_{1}^{Ana} x_{ij}+\REyAna_j+\epsilon^{Ana}_{ij}
    \quad\hbox{with }\epsilon^{Ana}_{ij}\iid \hbox{normal}(0,(\sigma_y^{Ana})^2)
    \end{equation} where the superscript, ``$Ana$" denotes the model for analysis or estimation as contrasted with the $DG$ model used for population data generation.
    We are interested in estimating $\beta_0^{Ana}$ and the standard deviation of the 
    PSU-REs, $\sigma_{\REyAna}$ where $\REyAna_j\iid \hbox{normal}(0,(\sigma_{\REyAna})^2)$. 
    (Note that the estimation of $\beta_1^{Ana}$ is unbiased regardless of the sampling scheme)
    We, subsequently, use the conditional estimation model for the marginal inclusion probabilities of \eqref{eq:lonnormalpriorforpi} to include a PSU-indexed random effects term,
    \begin{equation*}
    \log \pi_{ij}\mid y_{ij},\REpipop_j\sim 
    \hbox{normal}(\kappa_0+\kappa_y y_{ij}+\kappa_x x_{ij}+\REpipop_j,\sigma_\pi^2)
    \end{equation*}
    
    These two formulations describe the joint population model that employs random effects in both the marginal model for the response and conditional model for the inclusion probabilities.  We leverage the Bayes rule approach of Section~\ref{subsec:reviewLN} under the linear regression population to produce \eqref{eq:LRp_s} that adjusts the population model to condition on the observed sample.  We use this equation to estimate the Fully Bayes population model on the observed sample.
    
    It bears noting that FULL.both assumes that the data analyst does not have access to PSU-indexed sampling weights ($\propto 1/\pi_{1j}$).  Yet, we show in the simulation study results that FULL.both is able to adjust for informative sampling of PSUs for estimation of population model parameters (\eg, the intercept and PSU random effects variance).  This relatively good result owes to the inclusion of PSU-indexed random effects in the conditional model for the inclusion probabilities, $\pi_{ij}$, because it captures the within PSU dependence among them.
    
    Note that the Full.both analysis assumes that  $\log \pi_{ij}\mid y_{ij},\dots$ is a normal distribution, but that does not hold under any simulation scenario, which allows our assessment of the robustness of FULL.both to model misspecification.
    \item \textbf{FULL.y}: This alternative is a variation of FULL.both that uses the same population estimation for the response stated in ~\eqref{eq:AnalysisModel}.  In this option, however, PSU-REs are \emph{excluded} from the conditional model for the marginal inclusion probabilities; \ie,
    $\log \pi_{ij}\mid y_{ij},\REpipop_j\sim 
    \hbox{normal}(\kappa_0+\kappa_y y_{ij}+\kappa_x x_{ij},\sigma_\pi^2)$
    does not include PSU-REs.

    %\item \textbf{Pseudo.w}: Denotes the pseudolikelihood of Section~\ref{subsec:pseudo} where the likelihood is exponentiated by marginal sampling weights for individuals and the prior for the random effects is expontentiated with PSU-indexed sampling weights. In \eqref{eq:fullpseudo} we use $\sampled{w}_{1j}\propto1/\sampled{\pi}_{1j}$ (and not $\sampled{w}_{\cdot j}$) standardized so that $\sum_{j=1}^J \sampled{w}_{1j}=n$  
    \item \textbf{Pseudo}:  Denotes the pseudolikelihood that exponentiates the likelihood by marginal sampling weights, 
    %but does \emph{not} exponentiate the prior distribution for random effects by PSU-indexed sampling weights, 
    as described in Subsection ~\ref{subsec:pseudo}. 

    \item \textbf{Freq}: Denotes the frequentist, design-based, inference under simple linear regression model as described in Subsection 
    \ref{subsec:freq}.
    %$y_{ij}\sim \hbox{Normal}(\beta_{0}^{Ana}+\beta_{1}^{Ana}(\sigma_y^{Ana})^2)$.
    Note that this analysis model does not include PSU- REs because we employ a step that resamples the PSUs in order to estimate confidence intervals.  To fit the model, we use R function svyglm in library survey~\citep{lumley2019surveyR}.
    %(the sampling design passed to the function has one stratum with $J=30$ PSUs.)
    \item \textbf{Pop}: Ignore the informative sampling  and fit model in  \eqref{eq:AnalysisModel}
    (as if the sample were a cSRS).
    The inclusion probabilities do not play a roll in the inference, though the model for the response \emph{includes} PSU-REs. This is equivalent to Pseudo with 
    all %(group and individual) 
    sampling weights set equal to 1.

\end{enumerate}
\item Analyze the cluster simple random sample. \textbf{cSRS}: Fit the model in \eqref{eq:AnalysisModel} to the sample taken under a cSRS (generated in step 6) design. The same as Pop but applied to the cSRS generated in step 6.

\item Save parameter estimates to compute Bias, MSE, coverage probability of central 95\% credible intervals and their expected length. The parameters of inferential interest are the point estimate of $\beta_0^{Ana,TRUE}$: 
$\tilde{\beta}_0^{Ana}:=E(\beta_0^{Ana}\mid data)$ 
(or $\tilde{\beta}_{0,freq}^{Ana}$ for Freq), and its  central 95\% credible (or confidence for Freq) interval lower and upper limits.
We also produce point and interval estimates for $\sigma_y^{Ana,TRUE}$ for those methods that include PSU-REs in the marginal response model
(\ie\ exclude Freq). The computation of 
$\beta_0^{Ana,TRUE}$ and $\sigma_y^{Ana,TRUE}$ is discussed in Subsection
\ref{subsec:TrueModelParameters}.
%Of both $\beta_{0,Analysis}^{TRUE}$ and $\beta_{1,Analysis}^{TRUE}$. We will mainly focus on the estimation of $\beta_{1,Analysis}^{TRUE}$ (that is equal to 1).
%\item For computational reasons we can generate the population data fixed across all the simulations. That is perform steps 1-7 only once; 8 and 10 $M=10^3$ (say) times and use those estimates to compute the quantities in 11.
%\item Analyze the cSRS  using the model for the response ()

\end{enumerate}
Once we have run steps 1-10  $1000$ times, 
we use the quantities stored in step 10 to estimate the bias and MSE of the points estimate 
of $\beta_0^{Ana,TRUE}$ (defined below in Subsection \ref{subsec:TrueModelParameters}) 
of each method as the average of $\tilde{\beta}^{Ana}_0-\beta^{Ana,TRUE}_0$ and average of 
$(\tilde{\beta}^{Ana}_0-\beta^{Ana,TRUE}_0)^2$, respectively.
The coverage and expected length of the 95\% credible (or confidence) are estimated as the
proportion of times that the credible intervals contain  $\beta^{Ana,TRUE}_0$ and their
average length. We do the same for $\sigma_{\REyAna}^{TRUE}$ also defined in subsection 
\ref{subsec:TrueModelParameters}.

\subsection{True Model Parameters under Analysis Model}\label{subsec:TrueModelParameters}
In this section we compute the true values of the intercept and random effects variance parameters for the analysis ($Ana$) models that are obtained from associating parameters of the analysis model to the data generating ($DG$) model.  Having true values for the intercept and random effect variance under our analysis models allows our assessment of bias, MSE and coverage. We use the superindex ``$TRUE$" to refer to the true parameter values for the $Ana$ model implied by the simulation true parameter values in the $DG$ model. The true value of the intercept parameter under the analysis model is achieved by integration,
 $$\beta_0^{Ana,TRUE}=E(y_{ij}\mid x_{ij}=0)=
\beta_0^{DG}+\beta_{\pi,1} E(\pi_{1j})+\beta_{\pi,2} E(\pi_{i\mid j})
%+\beta_{\REyDG}E(\pi_{1j})\, \underbrace{E(\REyDG)}_{=0},
$$
yielding, $\beta_0^{Ana,TRUE}=2,1$ and $0$  under simulation scenarios \Sai, \Sc\ and \Sd, respectively. The true value for the population random effect is, 
${\REyAna_j}^{,TRUE}=\beta_{\pi,1} \left[\pi_{1j}-E(\pi_{1j})\right]+
                \beta_{\REyDG} \REyDG_j
                %+\beta_{\pi,\REyDG} (\pi_{1j} \times \REyDG_j)
                $  
and the true values random errors under the analysis model are 
$\epsilon_{ij}^{Ana,TRUE}=\beta_{\pi,2} \pi_{i\mid j}+\epsilon_{ij}^{DG}$.
Since $\pi_{i\mid j}$s are not normally distributed, $\epsilon_{ij}^{Ana,TRUE}$s are also not normally distributed. %, which as earlier mentioned allows our assessment of the robustness of the FULL.both analysis model to misspecification.  
Since the normality assumption of the errors of the simple regression model is violated, the variance of random effects for the marginal population model for the response, 
$$
\begin{array}{rl}
\Var({\REyAna_j}^{TRUE})=&\beta_{\pi,1}^2 \Var(\pi_{1j})+
\beta_{\REyDG}^2 \underbrace{\Var(\REyDG_j)}_{\sigma_{\REyDG}^2}
%+\beta_{\pi,\REyDG}^2 var(\pi_{1j}) \, var (\REyDG_j)\\
%-&
%2 \beta_{\pi,\REyDG} \beta_{\REyDG} E(\pi_{1j})\, var (\REyDG_j)
\end{array}
$$
is different from $(\sigma_{\REyAna}^{TRUE})^2$. 
Nevertheless, we may compute $\sigma_{\REyAna}^{TRUE}$ by fitting the 
linear mixed effect population model in \eqref{eq:AnalysisModel}, which corresponds to the analysis model for the response,  directly to the population dataset via the \text{lmer} R function
%n the simulation study, R program “simulationswithISWOreplacement.R”, insteadof using (1) to estimate the value ofσTRUEη,analysisI estimated this parameter via Monte-carlo with R function 
lmer($y\sim x + (1\mid$ PSU index),data= population).  In practice, the PSU inclusion probabilities, $(\pi_{1j})$, are not available to the data analyst 
for either sampled or non-sampled individuals from the population. 
%\footnote{Luis 2/26/2020: Funny that lmer estimates 
%of the variance of the PSU sepecific RE $(\sigma_y^{Ana})^2\approx (\sigma_y^{DG})^2+\beta_{\pi,2}^2 var(\pi_{ij})$ 
%but the lmer-variance of the RE is not close to
%$var(\REyAna)$ in the equation\\ T any comments on that?}
Under scenario \Sai\ 
$\sigma_{\REyAna}^{TRUE}\approx 
0.27$, and %Scenario ai
under the other two scenarios $\sigma_{\REyAna}^{TRUE}\approx 0.1$.
%scenario d
%Notice that under scenario a(i), where  
%$\sigma_{\REyDG}^{TRUE}$=0 while 
%$\sigma_{\REyAna}^{TRUE}=0.27>0$.  %0.2699188$

%%%
%%Relabel the PSUs so $j$ runs from $1,\dots,J$. Call $\sampled{\pi}_{1,j}$ the sampled ${\pi_{1,j}}$s 

%Generate $S\subset \{1,2,\dots,N\}$, the set of indices in the sample.
%\begin{enumerate}
%\item For $j=1,\dots, J$, generate $S_j\subset \{1,2,\dots,N_{total}\}$  by sampling  
%$n_j=N_j/100$ indices with  $Pr(i\in S_j)=\pi_{i\mid j}/\sum_{\{i^\prime\in\hbox{\tiny{PSU }} j\}}\pi_{i^\prime\mid j}$.
%\item Relabel the $y_i$s so $i$ runs from $1,2,\dots,n$ with $n=\sum_{j=1}^J n_j$.
%So we end up with $\{\pi_{1j},\pi_{i\mid j},\pi_{i}\}_{i=1,\dots,n}$
%\end{enumerate}

%In all these scenarios also get a SRS just by replacing $\sampled{\pi}_{\star}$ with ${\pi}_{\star}^{srs}$  where $\star\in\{1j,i\mid j,i\}$ accordingly.

\subsection{Simulation Results}\label{subsection:simresults}
Tables \ref{tab:informativePSURE}, \ref{tab:Non-informativePSU-RE} and \ref{tab:nostageinformative}
show the simulation results under all scenarios. 
As expected in all scenarios cSRS credible intervals have coverage close
to nominal level (0.95) and the lowest MSE.
In informative scenarios, \Sai\ and \Sc,
(i) Pop performs poorly showing
the consequences of ignoring the informative sampling scheme,
(ii) all methods to analyze informative samples yield similar quality point estimators (similar MSE),
and 
(iii) FULL.both and FULL.y credible intervals maintain nominal coverage while %Pseudo.w, 
Pseudo and Freq do not.   
Under non-informative scenario \Sd, 
all methods to analyze informative samples
yield similar results to Pop (now correctly specified model). % since under this scenario $\pi_{ij}\perp y_{ij}$). 
Both Pseudo and Freq under-estimate uncertainty such that they both under cover in the informative sampling case.  Interestingly, only Freq pays the price of the noise introduced by the non-informative sampling weights
producing considerable wider confidence intervals for $\beta_0^{Ana,TRUE}$ than all other methods.

Overall, Tables \ref{tab:informativePSURE}-\ref{tab:nostageinformative} 
show that FULL.both and FULL.y are the best methods
to analyze informative samples, particularly in terms of uncertainty quantification.
But, so far, results have not shown advantage of FULL.both over FULL.y.
%it is not clear the need of having a PSU-RE on the model for $\pi_{ij}$.
%We decrease value of $a_\pi$ in 
%the distribution of $\pi_{i\mid j}\iid \hbox{Gamma}(a_\pi,2)$ in point 1 in Subsection 
%\ref{subsec:MCscheme}. This decrease the variance 
To do so, under scenario \Sai, we increase level of informativeness of the PSUs by increasing the value of
$\beta_{\pi,1}$ (See step 5 in Subsection \ref{subsec:MCscheme}) from
$J_{pop}$ to $2J_{pop}$ and $3J_{pop}$. 
As shown in Table \ref{tab:coverage}, 
coverage of FULL.y credible intervals deteriorates as informativeness increases 
while FULL.both, in contrast to all other methods, 
maintains coverage 
similar to cSRS  (at nominal level).

The strength of FULL.both over all other 
considered approaches  is that it accounts for the association among the inclusion probabilities within the same PSU.
Table \ref{tab:coverage} shows that
FULL.both is the only approach that performs well under simulation scenario \Sai\  
when the level of informativeness  of $\pi_{ij}$ (or correlation between $y_{ij}$ and $\pi_{1j}$) increases. 
FULL.both is 
the only method whose inference quality is not affected
when
%robust against violation of its assumption 
$\pi_{1j}\not\perp y_{ij}\mid \bx_{ij} \forall i,j$.
Since the  population simulation true
distribution of $\pi_{ij}$, given in 
point 7 (3.) in
Subsection \ref{subsec:MCscheme},
is not lognormal the simulation shows that FULL.both is robust to
misspecification of the distribution of 
$\pi_{ij}\mid y_{ij},\cdots$.
\ech

% latex table generated in R 3.5.2 by xtable 1.8-3 package
% Mon Oct 21 10:03:23 2019
% Table generated by Program simulations_with_InformativeREandFreq.R
% Processing output of simulations_with_ISWOreplacement.R
\begin{table}[ht]
\centering
\begin{small}
\begin{tabular}{rrrr  rrr}
                        & FULL.both     & FULL.y & Pseudo & Freq & Pop & cSRS\\ 
 \hline
 & \multicolumn{6}{c}{$\beta_0^{Ana}$}\\
 Bias                   & 0.035 & 0.046 & 0.067  & 0.015 & 0.518 & 0.005 \\ 
 MSE                    & 0.028 & 0.028 & 0.028  & 0.030 & 0.294 & 0.016 \\ 
 95\% CI Coverage\% CI & 0.949 & 0.948 & 0.902 &  0.905 & 0.088 & 0.957 \\ 
 95\% CI Length 95\% CI & 0.670 & 0.668 & 0.538 & 0.617 & 0.620 & 0.520 \\ 
\hline
 & \multicolumn{6}{c}{$\sigma_{\REyAna}$}\\
 Bias & 0.012 & 0.012 & 0.043               &NA   & 0.014 & -0.012 \\ 
  MSE & 0.010 & 0.010 & 0.010               &NA   & 0.010 & 0.008 \\ 
  95\% CI Coverage & 0.964 & 0.967 & 0.902  &NA  & 0.961 & 0.951 \\ 
  95\% CI Length  & 0.424 & 0.422 & 0.360   &NA   & 0.416 & 0.357 \\ 
  \hline
\end{tabular}
\end{small}
\caption{Simulation Scenario \Sai: Informative PSU-RE.
  cSRS analyses the cSRS sample while all other approaches analyze the informative sample. 
  CI denotes central credible interval except for Freq where it denotes confidence interval.
  NA stands for not applicable, Freq does not include PSU-REs.
  %$\sigma_{\REyAna}^{TRUE}\approx 0.27$
  %Notice that $\sigma_\gamma$ in the generating model is zero but $\sigma_\eta$ in the analysis model is not.
  \label{tab:informativePSURE}
  } 
\end{table}
 %Mon Oct 28 
% Table generated by Program 
%C:\Users\Luis Leon-Novelo\Google Drive\projects\Bayesian_survey_extension\results\informativeRE_withFREQ\only2ndstageinformative_table17
% Processing output of simulations_with_InformativeREandFreq
%  tABLE 17 IN BAYESIAN_SURVEY_EXTENSION overloaf file.
\begin{table}[ht]
\centering
\begin{small}
\begin{tabular}{rrrrrrr}
                & FULL.both & FULL.y      & Pseudo & Freq & Pop   & cSRS\\ 
  \hline
&  \multicolumn{6}{c}{$\beta_0^{Ana}$}\\
    Bias        & 0.008     & 0.009      & 0.055 & 0.011 & 0.494 & 0.001\\ 
    MSE         & 0.020     & 0.020     & 0.022 & 0.025 & 0.263 & 0.014\\ 
95\% CI Coverage& 0.958    & 0.962      & 0.908 & 0.926 & 0.064 & 0.962\\ 
95\% CI Length  & 0.623    & 0.624      & 0.496 & 0.565 & 0.574 & 0.479\\  
\hline
 & \multicolumn{6}{c}{$\sigma_{\REyAna}$}\\
Bias            & 0.063     & 0.065          & 0.107 &NA       & 0.065& 0.049 \\ 
  MSE           & 0.008     & 0.008          & 0.017 &NA       & 0.008 & 0.006\\ 
95\% CI Coverage& 0.967     & 0.971          & 0.752 &NA       & 0.966 & 0.956\\ 
95\% CI  Length & 0.332     & 0.331          & 0.312 &NA       & 0.329& 0.285 \\ 
\hline
\end{tabular}
\end{small}
\caption{Simulation Scenario \Sc: Non informative PSU-REs. %(or only2ndstageinformative)
Same as Table \ref{tab:informativePSURE} but under \Sc.
%cSRS analyses cSRS samples while all other approaches analyze  informative samples.
\label{tab:Non-informativePSU-RE}
 %$\sigma_{\REyAna}^{TRUE}\approx 0.1$
  } 
\end{table}

% Mon Oct 28 
% Table generated by Program analyzing_simulations_with_ISWOreplacement.R
% Processing output of simulations_with_InformativeREandFreq.R
%  tABLE 18 IN BAYESIAN_SURVEY_EXTENSION overloaf file.
\begin{table}[ht]
\centering
\begin{small}
\begin{tabular}{rrrrrrr}
& FULL.both & FULL.y  & Pseudo & Freq& Pop & cSRS\\ 
  \hline
 & \multicolumn{6}{c}{$\beta_0^{Ana}$}\\
Bias                & 0.000 & 0.000  & -0.000 & 0.000 & -0.000 & 0.000 \\ 
  MSE               & 0.001 & 0.001  & 0.001 & 0.001 & 0.001 & 0.001 \\ 
95\% CI   Coverage  & 0.957 & 0.964  & 0.944 & 0.947 & 0.956 & 0.951 \\ 
95\% CI   Length    & 0.103 & 0.103  & 0.104 & 0.134 & 0.102 & 0.103 \\ 
  \hline
 & \multicolumn{6}{c}{$\sigma_{\REyAna}$}\\
Bias                & 0.002 & 0.002  & 0.006 &NA  & 0.002 & 0.003 \\ 
  MSE               & 0.000 & 0.000  & 0.000 &NA  & 0.000 & 0.000 \\ 
95\% CI   Coverage  & 0.935 & 0.936  & 0.933 &NA  & 0.938 & 0.955 \\ 
95\% CI   Length    & 0.067 & 0.067  & 0.068 &NA  & 0.067 & 0.068 \\ \hline
\end{tabular}
\end{small}
\caption{Simulation Scenario \Sd: No stage informative.
Same as Table \ref{tab:informativePSURE} but under \Sd,\ where Pop is correctly specified. 
%cSRS analyses cSRS samples while all other approaches do the same as if the samples were informative.
 %tRUE VALUE  truesigma_eta_analysismodel=0.1004758 
  \label{tab:nostageinformative}
  } 
\end{table}

%\afterpage{\FloatBarrier}

\begin{table} [!htb] 
\begin{tabular}{ l | cccccccc}
$\beta_{\pi,1}$&FULL.both   & FULL.y      & Pseudo        &Freq           &cSRS   \\
\hline
 $J_{pop}$      &.949,.964  &.948,.967   &.902,.902       & .905,NA       &.957,.951  \\ %tABLE 11 IN BAYESIAN_SURVEY_EXTENSION
 $2J_{pop}$     &.949,.929  &.914,.927    &.852,.929      & .908,NA       &.961,.927\\  %TABLE 16
 $3J_{pop}$     &.949,.946  &.85,.954     &.808,.914      & .910,NA       &.957,.949\\  %TABLE 22       
\end{tabular}
\caption{
Coverage of central 95\% credible (confidence for Freq) intervals for
$\beta_0^{Ana,TRUE}$,$\sigma_{\REyAna}^{TRUE}$ 
under scenario 
\Sai\ increasing the level of informativeness of the PSUs (by increasing $\beta_{\pi,1})$.
NA stands for not applicable, Freq does not include PSU-REs.
\label{tab:coverage}
%Results summary $\checkmark$   $\approx 0.95$ coverage; 
%$\times$ (coverage for $\beta_0,\sigma_\eta) \neq .95$  coverage.\\
%$a_\pi=2$  low variance on $\pi_{1j}$, 
%$a_\pi=1$  high variance on $\pi_{1j}$, 
%$\beta_{\pi,1}=2J_{pop}$ higher association between RE and response.
}
\end{table}
%
%\afterpage{{\FloatBarrier}} 
\section{Application}\label{sec:applications}
The National Health and Nutrition Examination Survey (NHANES) is  designed
to assess the health and nutritional status of the non-institutionalized  civilian population living in one of the 50 U.S. states and Washington D.C.  
Although nationally representative, NHANES is designed to oversample specific subpopulations 
(\eg\  persons 60 and older, African Americans, Asians, and Hispanics)
and follows a complex sampling design \citep{CDCG}.
The NHANES sampling design is constructed as multi-stage  with stages that include sampling strata and nested primary sampling units (PSUs) that further nest respondents. A PSU is a cluster or grouping of spatially contiguous counties, while a stratum is  a region nesting multiple PSUs.  
NHANES publishes respondent-level marginal sampling weights based on resulting respondent marginal inclusion probabilities in the sample after accounting for clustering. 
\begin{comment}
The sampling weights measure the number of people in the population represented by that sampled individual, 
reflecting unequal probability of selection, nonresponse adjustment, and adjustment to independent population controls. 
\end{comment}

The survey consists of both interviews and physical examinations. 
The NHANES interview includes demographic, socioeconomic, dietary, and health-related questions. The examination component consists of medical, dental, and physiological measurements, as well as laboratory tests.
Data obtained from $J=30$ PSUs, corresponding to 
15 strata with two PSU per stratum, are released in two-year cycles.
The example considers the dietary data.
The analyses consider PSU information and sampling weights as provided by NHANES but does not incorporate strata information.

In this application, we estimate the average kilocalories (kcal) consumed in each one of the gender, 
age and ethnicity groupings.
We use dietary data from the 2015-2016 NHANES cycle.  
Each participant  answers a 24-hour dietary recall interview in two days: Day 1 and Day 2.
The Day 1 recall interview takes place when the participant visits the Mobile Exam Center  (MEC) unit where other NHANES measurements are taken. The Day 2 recall interview is collected by telephone and it is scheduled for 3 to 10 days later (See \cite{CDCF} for more details).
Based on theses interviews NHANES provides datasets with estimates 
of kilocalories (and many nutrients)  ingested by the participant 24 hours before the interview along with their dietary sampling weight. In this application, we consider the Day 1 dataset and the sampling weights that come in it. 
%See Subsection \ref{subsec:kilocalappdetails} for more details.
%as used in the previous application 
%In this cycle 15,327 persons were selected; of these 
%In this cycle 9,544 were considered participants to the MEC examination and data collection. 
%In this cycle a total of 8,506 MEC participants provided complete dietary intakes for Day 1, and of those providing the Day 1 data, 
%7,027 provided complete dietary intakes for Day 2.  

There are 8,506 participants who completed the Day 1 dietary recall, of which this analysis
considers the $n=8,327$ with positive sampling weights or, equivalently, with recall status labeled ``complete and reliable'' by NHANES \cite{CDCF}. 
%We fitted the same multiple linear regression models as in the previous example with response  $y=\log(\hbox{kcal}+1)$ with kcal the NHANES estimate of kilocalories consumption based on Day 1 recall interview.

The underlying analysis model for the non-frequentist methods 
(FULL.both, FULL.y, Pseudo and Pop) is the 
mixed effect linear regression with response
 $y=\log(\hbox{kcal}+1)$ with kcal the NHANES estimate of kilocalories consumption based on Day 1 recall interview; predictors: gender, age group and race/ethnicity; and, PSU-REs.
The frequentist analysis model, Freq, is the same (now fixed effect) model but without PSU-REs.
Age is  categorized in 5 groups: $[0,8],[9,17],[18,29],[30,49]$ and $[50,80]$ years old, while race/ethnicity categories are 
 non-Hispanic White,
 Mexican American, 
 other-Hispanic,
 non-Hispanic Black, and 
 other or multiracial.
 Male, $[0,8]$ age group and non-Hispanic White are the reference groups. 
We recall that $\bxy$ denotes predictors in the marginal model for $y$ in   \eqref{eq:SLR_likelihood}, and construct,
\begin{equation}%\label{eq:bxyinfirstapplication}
\begin{array}{rl}
\bxy^t=\big(&1,1(gender=\hbox{Female}),\\
&1(Age\in [9,17]),1(Age\in [18,29]),
1(Age\in [30,49]),1(Age\in [50,80]),\\
&1(Race/Eth=\hbox{Mexican American}),1(Race/Eth=\hbox{other Hispanic}),\\
&1(Race/Eth=\hbox{non-Hispanic Black}),
1(Race/Eth=\hbox{other or multiracial})\big)
\end{array}
\end{equation}
with dimension $p+1=10$, where $1(A)$ denotes the indicator function of the individual in the set $A$.   
In this application, we set $\bxp=\bxy$
 in  \eqref{eq:lonnormalpriorforpi}.
%In the implementation of Pseudo.w we use sampling weights, $\sampled{w}_{\cdot j}$ of \eqref{eq:fullpseudo}, that sum individual sampling weights for those units nested in each PSU in order to exponentiate the random effects prior distribution.

For the non-frequentist methods, 
priors are assigned in \eqref{eq:priors}, and 
posterior inference 
is based on a posterior sample 
of the model parameters of size 10,000. 
The MCMC sampler was run 10,000 iterations, after a burn-in period of another 10,000 iterations, on Stan.  Relatively fewer draws are required when using Stan's Hamiltonian Monte Carlo (HMC) than a Gibbs sampler because the Stan draws are less correlated.

\begin{figure}
%Data preprocessing: Application_Cupper.R 
%Data analysis: Application_FATvsBMI.R
%Workspace Kilocal20200516.RData in line 685 of program above (after the Gibbs sampler have been performed
\begin{center}
%\begin{tabular}{rl}
%\includegraphics[width=80mm,angle=0]{\plotpath/beta0dr1tkcal.pdf}&
%\includegraphics[width=80mm,angle=0]{\plotpath/beta014dr1tkcal.pdf}&
%\end{tabular}
\includegraphics[width=120mm,angle=0]{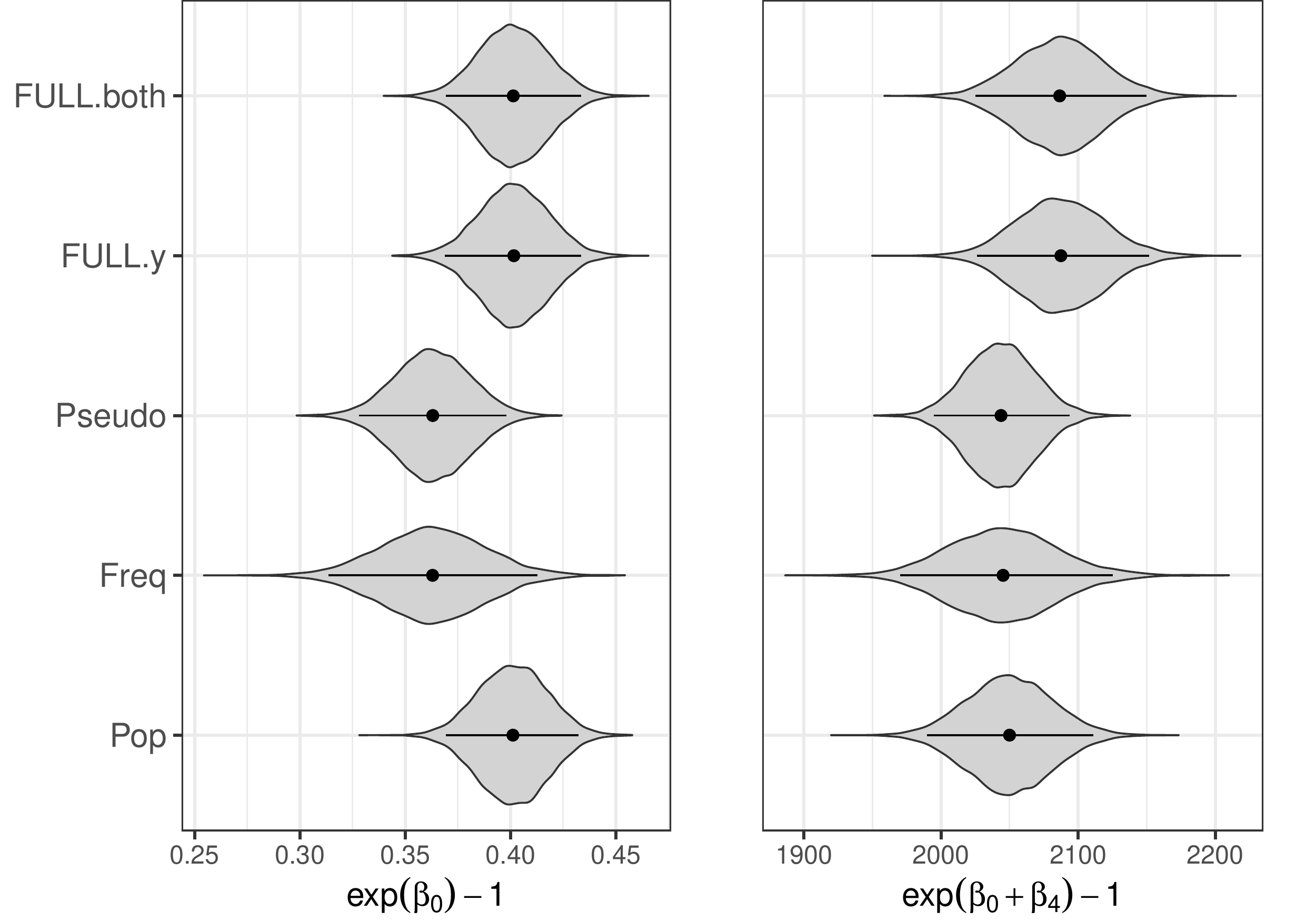}
\end{center}
\caption{\label{fig:beta0dr1tkcal} 
Violin plots, under all methods, for average kcal consumption for people in the reference group 
White males 8 year old or younger (left), and White males in the  age group [30,49] (right)
 along with point estimate (dot) and central 95\% credible, or confidence, interval (horizontal line within violin plot). 
}
\end{figure}

Figure \ref{fig:beta0dr1tkcal} depicts violin plots of the estimated mean daily kcal consumption
for White males in age groups [0,8] (left) and [30,49] (right).
More specifically, the left panel depicts 
violin plots  
of the posterior distribution of $\exp(\beta_0)-1$ for the set of non-frequentist methods.  %(FULL.both, FULL.y, Pseudo and Pop). %(along with posterior mean and central 95\% CI); 
%and,  
%with the posterior mean point and distribution estimates for $\beta_0$ drawn from the
%posterior distribution under the  Bayesian methods.  
For the frequentist method (Freq) depicts
the distribution of $\exp(\beta_0)-1$ with $\beta_0$ drawn from   
$\hat{\beta}_0+t\times SE(\hat{\beta}_0)$ with $t\sim \hbox{Student-t}$ with 
$J-1=30-1=29 $ degrees of freedom.
The right panel depicts the violin plot of the
posterior distributions of 
$\exp(\beta_0+\beta_4)-1$ for the non-frequentist methods.  It also depicts 
the violin plot of $\exp(\beta_0+\beta_4)-1$ for Freq, though with 
$(\beta_0,\beta_4)$  drawn from the distribution 
$(\hat{\beta}_0,\hat{\beta}_4)^t+ \hat{\Sigma}_{1,4}^{1/2} \mathbf{t}$ where the 
random vector
$\mathbf{t}=(t_0,t_4)^t$ has entries $t_0,t_4 \iid \hbox{Student-t}$, 
with 29 degrees of freedom %the same degrees of freedom as above (\ie, 20),
and 
$\hat{\Sigma}_{0,4}^{1/2} \hat{\Sigma}_{0,4}^{1/2} =\hat{\Sigma}_{0,4}$
with $\hat{\Sigma}_{0,4}$ the estimated variance-covariance matrix of $(\hat{\beta}_0,\hat{\beta}_4)^t$.

Figure \ref{fig:beta0dr1tkcal} shows that for these groups, inference under FULL.both and FULL.y are similar to one another but different from Pop.
The FULL.both central 95\% credible interval  
(and also the FULL.y one, not shown) for $\kappa_y$, 
$(-0.110,-0.048)$, does not contain zero, indicating that the sampling design is informative.  FULL.both 
and FULL.y correct for this.

The point estimates under %Pseudo.w, 
Pseudo  and Freq are close to one another, but differ from those for FULL.both and FULL.y, indicating that the weight smoothing provided by the fully Bayesian methods is more robust to noise present in the weights that may injure point estimation.

\begin{comment}
Table \ref{table:Fullboth_kcal} displays inference for the model parameters under Full.Both.

Table \ref{table:Fullboth_kcal} confirms the expected pattern of kcal 
consumption; it
increases with age when young, plateau at middle age and decreases in the oldest age group. Table \ref{table:Fullboth_kcal} also shows that,
in average, White people consumes more kcals than each other race/ethnicity groups.\footnote{T: maybe White people are taller. We could have controlled for height, when controlling for BMI the design becomes non-informative.}
In contrast,  Freq.strata  
(See Table \ref{table:Freq.strata_kcal} in the appendix subsection \ref{subsec:kilocalappdetails})
concludes that the only group with, statistically significant 
(p-value<0.05)
lower kcal consumption than 
the White people group is the non-Hispanic Black people group.
\end{comment}

\begin{figure}
%Data preprocessing: Application_Cupper.R 
%Data analysis: Application_FATvsBMI.R
%Workspace Kilocal20200516.RData in line 685 of program above (after the Gibbs sampler have been performed
\begin{center}
\includegraphics[width=115mm,angle=0]{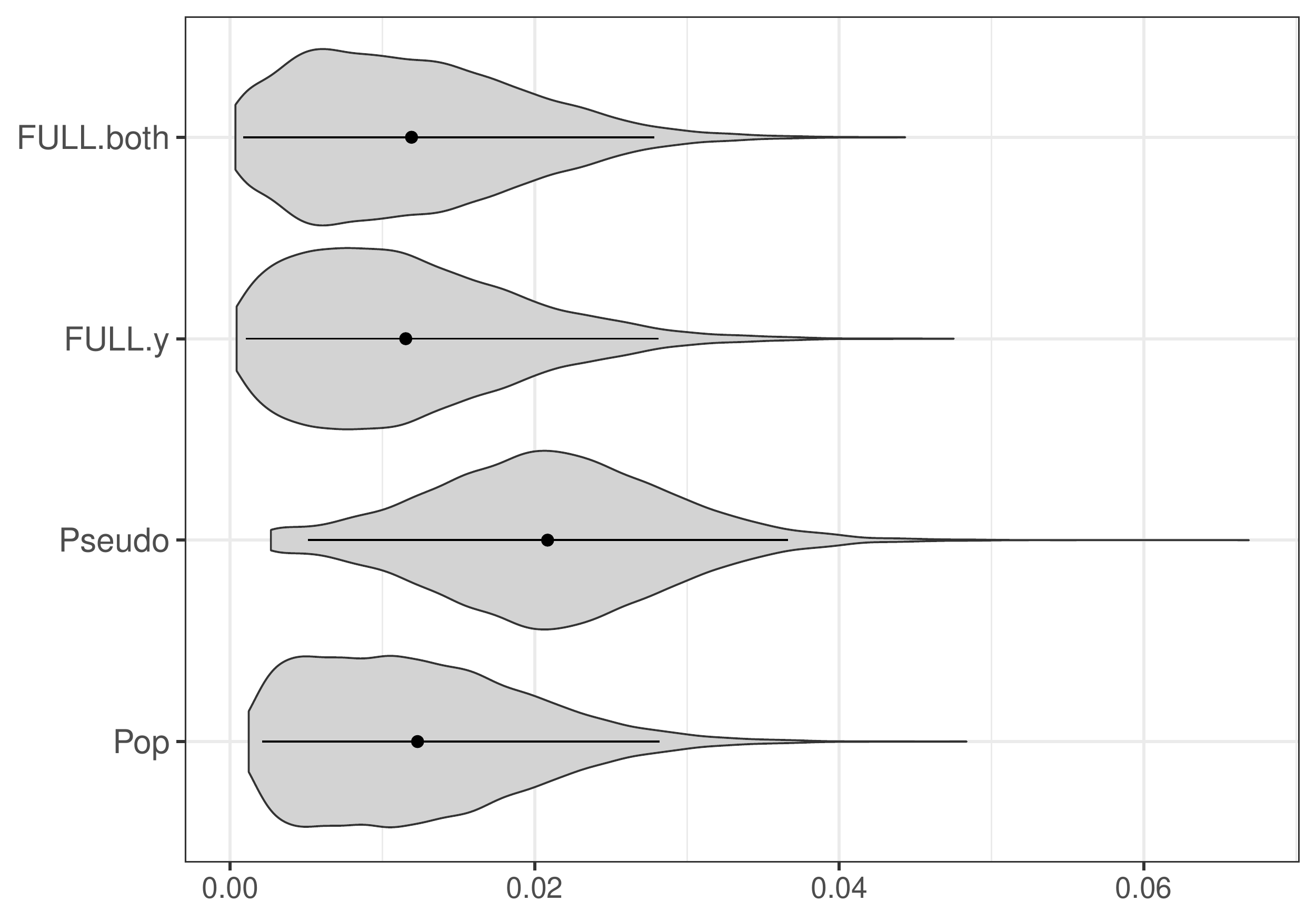}\\
\vspace{-3mm} $\sigma_{\REypop}$\\
\end{center}
\caption{\label{fig:sigma_ydr1tkcal} 
Violin plots of the estimate of the standard deviation of the PSU-RE, $\sigma_{\REypop}$,
under all non-frequentist methods.  
}
\end{figure}

\begin{figure}
%Data preprocessing: Application_Cupper.R 
%Data analysis: Application_FATvsBMI.R
%Workspace Kilocal20200516.RData in line 685 of program above (after the Gibbs sampler have been performed
\begin{center}
\includegraphics[width=120mm,angle=0]{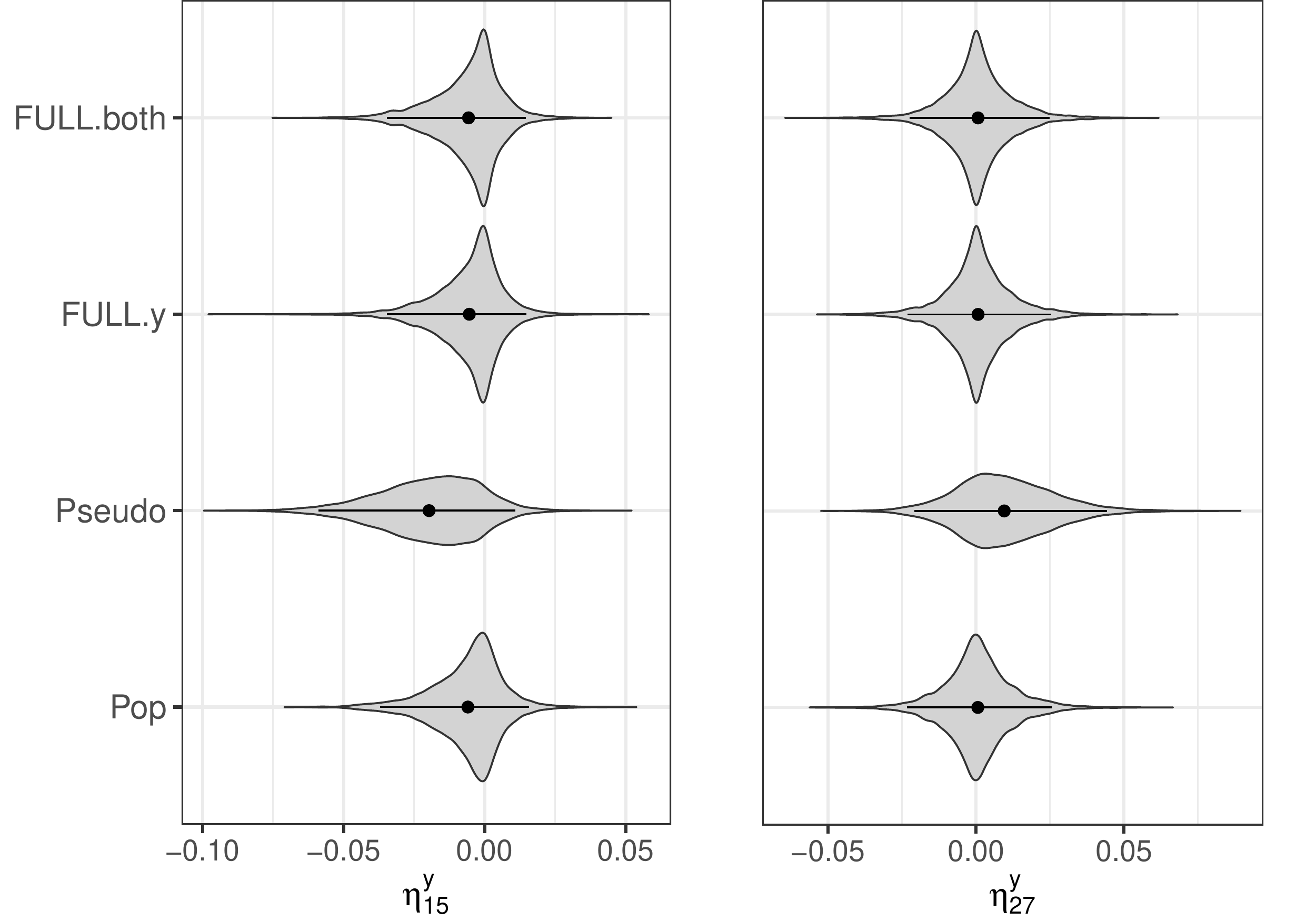}
\end{center}
\caption{\label{fig:PSUREinresponse15_27dr1tkcal} 
Violin plots, under all non-frequentist methods, 
for PSU-REs $\REypop_{15}$ and $\REypop_{27}$.  PSUs 15 and 27 here correspond to
PSUs 1 in strata  8 and  14, respectively, in the 2015-2016 demographic NHANES dataset.
}
\end{figure}

Figure \ref{fig:sigma_ydr1tkcal} depicts the violin plots for the standard deviation of the PSU-RE, $\sigma_{\REypop}$ under the non-frequentist 
%Pop, as well as FULL.both, Full.y and Pseudo 
methods 
(Freq does not model PSU-RE). Inference for this parameter under Pseudo differs from those under the two fully Bayesian methods.  
Figure \ref{fig:PSUREinresponse15_27dr1tkcal} shows that the posterior distribution of individual PSU random effects in the marginal response model (PSU-RE) also differs. The figure focuses on two particular random effects, $\REypop_{15}$ and $\REypop_{27}$, that are coherent with the general 
pattern we see over the random effects; in particular, the fully Bayes methods express less estimation uncertainty than does Pseudo, indicating a greater estimation efficiency by jointly modeling the response and inclusion probabilities. 

{\FloatBarrier} 
\section{Discussion}
We have extended our work in \LS\ to include PSU information in a model-based, fully Bayesian analysis of informative samples.
The extension consists of replacing the fixed effect model by a mixed effect model that includes 
PSU/cluster-indexed random effects in the marginal model for $y$ and the conditional model for $\pi\mid y$ to capture dependence induced by the clustering structure.
%to mixed effects models allows us to achieve asymptotically correct uncertainty quantification with the simple step of including PSU/cluster-indexed random effects in the marginal model for y and conditional model for \pi|y.   None of the comparator methods achieve the same.   Our FB-RE approach is robust to noise in weights.
We have shown via simulation that our fully Bayesian approach yields correct uncertainty 
quantification, or equivalently CIs, with coverage close to their nominal level, including for the random effects variances.  Competing methods fail to do so in at least one simulation scenario. In particular, FULL.both is the only appropriate method, of all here considered,  
when the sample design is informative for the selection of PSUs.  The results in simulation scenario \Sd, where the design is not informative, revealed that the method is also robust to noise in weights.

Our fully Bayesian methods proposed here are mixed effect linear models that not only take into account the possible association 
of individuals within the same cluster but also, in contrast to the design-based frequentist methods, quantify this association; that is, the within PSU correlation can be estimated. We demonstrated our method with an NHANES dietary dataset whose sampling design includes stratification.
The next natural step of the method is to include strata information into the analysis.
%We demonstrated our method with %two applications analyzing NHANES data.
%In the first application, the comparative methods behave as in simulation scenario \Sd; that is, 
%the Bayesian methods adjust the inference for noninformative design yielding similar results to an analysis assuming  
%cSRS where sampling weights do not play a role. 
%By contrast, inference under competing methods 
%is affected by the noise induced from these sampling weights.  In the second application the design is informative and inference under Bayesian methods differs from other methods. 
%In our two applications, inclusion of the strata information into the frequentist analysis did not yield shorter confidence intervals as we expected (Freq.strata not better than Freq), indicating that perhaps strata are homogeneous (with respect to the response/ predictor relationship considered). 
This  application only analyzed data from Day 1 dietary questionnaire.
To analyze Day 1 and Day 2 data with one model we need to adapt our approach to repeated measures. This is another current line of research.

To implement our Bayesian method, we derived an \emph{exact} likelihood for the observed sample. 
In principle, this likelihood can also  be used for maximum likelihood estimation opening the door for model-based frequentist inference. 

Our approach requires the modeler to specify a distribution for $\pi_i\mid y_i,\cdots$.
Estimation requires the computation of an expected value, the denominator in \eqref{eq:IScorrectionPSU}.  We assume a lognormal conditional likelihood for the marginal inclusion probability, given the response, with linear relationship between the location parameter and the the response, both of which facilitate use of Theorem \ref{th:closeformPSU} to obtain a closed form for this expected value. Our simulation study showed that the Bayesian method is robust against misspecification of these assumptions. Future work is needed to ease conditions in Theorem \ref{th:closeformPSU}. 

To sum up, we have presented the first model-based Bayesian estimation approach that accounts for  both informative sampling within the individuals in the same PSU
and when the PSU is informative to produce correct uncertainty quantification.

\appendix

\section{
 Quantity between Brackets  in Augmented Pseudolikelihood in \eqref{eq:fullpseudo}
 Matches  Likelihood under Weighted Linear Regression 
  \label{subsec:pseudoandweightedreg}
 }
 The  weighted linear regression model is 
${y_{ij}\mid \bxy_{ij},\bth,\REypop_j}\sim\text{normal}\left(\bxy_{ij}^t\bbe+\REypop_j ,\sigma_y^2/w_{ij}\right)$ with $w_{ij}>0$ known. 
 %Pseudolikelihhood estimation in Subsection \ref{subsec:pseudo}
%under  the regression model in \eqref{eq:SLR_likelihood}
%matches likelihood  under
Using the fact that  
$$\left[\text{normal}(y\mid \mu,\sigma^2)\right]^w= \frac{1}{w^{1/2}(2\pi)^{(w-1)/2}} \frac{1}{(\sigma^2)^{(w-1)/2}} \times \text{normal}(y\mid \mu,\sigma^2/w)
$$
%Setting $\mu_{ij}=\bxy_{ij}^t\bbe+\REypop_j$ to ease notation 
we  obtain that the expression between brackets in \eqref{eq:fullpseudo},
$\prod_{j=1}^J \prod_{i=1}^{n_j} p(\sampled{y}_{ij}\mid \theta)^ {\sampled{w}_{ij}}$, equals

\begin{align*}
\prod_{j=1}^J 
\prod_{i=1}^{n_j} 
\left[\text{normal}\left(\sampled{y}_{ij}\mid \bxy_{ij}^t\bbe+\REypop_j ,\sigma_y^2 \right)
\right]^{\sampled{w}_{ij}} \propto& 
\frac{1}{(\sigma_y^2)^{\left(\sum_{j,i}[\sampled{w}_{ij}-1]\right)/2}}\\
&\times
\left[\prod_{j=1}^J 
\prod_{i=1}^{n_j} \text{normal}\left(\sampled{y}_{ij}\mid \bxy_{ij}^t\bbe+\REypop_j ,\sigma_y^2/\sampled{w}_{ij} \right)
\right]
\end{align*}
but, by construction, $\sum_{j=1}^J\sum_{i=1}^{n_j} \sampled{w}_{ij}=n$ and therefore the expontent of $\sigma_y^2$ in the denominator in the expression above is zero,  
and the quantity between brackets is the likelihood of the weighted linear regression model.
\bibhang=1.7pc
\bibsep=2pt
\fontsize{9}{14pt plus.8pt minus .6pt}\selectfont
\renewcommand\bibname{\large \bf References}
%\begin{thebibliography}{11}
\expandafter\ifx\csname
natexlab\endcsname\relax\def\natexlab#1{#1}\fi
\expandafter\ifx\csname url\endcsname\relax
  \def\url#1{\texttt{#1}}\fi
\expandafter\ifx\csname urlprefix\endcsname\relax\def\urlprefix{URL}\fi
%\fi

\bibliographystyle{ba}

\bibliography{mv_refs_sep2015_ss,SurveywWeights}

\begin{thebibliography}{13}
\newcommand{\enquote}[1]{``#1''}
\expandafter\ifx\csname natexlab\endcsname\relax\def\natexlab#1{#1}\fi
\expandafter\ifx\csname url\endcsname\relax
  \def\url#1{{\tt #1}}\fi
\expandafter\ifx\csname urlprefix\endcsname\relax\def\urlprefix{URL }\fi
\ifx\endbibitem\undefined \let\endbibitem\relax\fi

\bibitem[{Carpenter et~al.(2016)Carpenter, Gelman, Hoffman, Lee, Goodrich,
  Betancourt, Brubaker, Guo, Li, and Riddell}]{carpenter2016stan}
Carpenter, B., Gelman, A., Hoffman, M., Lee, D., Goodrich, B., Betancourt, M.,
  Brubaker, M.~A., Guo, J., Li, P., and Riddell, A. (2016).
\newblock \enquote{Stan: {A} probabilistic programming language.}
\newblock {\em Journal of Statistical Software\/}, 20: 1--37.
\endbibitem

\bibitem[{CDC(2011)}]{CDCNHANESSurveyDesign}
CDC (2011).
\newblock \enquote{Centers for {D}isease {C}ontrol ({CDC}). {C}ontinuous
  {NHANES} Web Tutorial: Key Concepts about {NHANES} Survey Design.}
\newblock Accessed: May 21, 2018.
\newline\urlprefix\url{https://www.cdc.gov/nchs/tutorials/NHANES/SurveyDesign/SampleDesign/Info1.htm}
\endbibitem

\bibitem[{CDC(2015)}]{CDCF}
--- (2015).
\newblock \enquote{Centers for {D}isease {C}ontrol ({CDC}). {{NHANES}:
  Measuring Guides for the Dietary Recall Interview}.}
\newblock Page last reviewed: November 6, 2015. Accessed: May 14, 2020.
\newline\urlprefix\url{https://www.cdc.gov/nchs/nhanes/measuring_guides_dri/measuringguides.htm}
\endbibitem

\bibitem[{CDC(2017)}]{CDCG}
--- (2017).
\newblock \enquote{Centers for {D}isease {C}ontrol ({CDC}). {{NHANES}: About
  the National Health and Nutrition Examination Survey}.}
\newblock Page last reviewed: September 15, 2017. Accessed: May 14, 2020.
\newline\urlprefix\url{https://www.cdc.gov/nchs/nhanes/about_nhanes.htm}
\endbibitem

\bibitem[{Kleijn and van~der Vaart(2012)}]{kleijn2012}
Kleijn, B. and van~der Vaart, A. (2012).
\newblock \enquote{The Bernstein-Von-Mises theorem under misspecification.}
\newblock {\em Electron. J. Statist.\/}, 6: 354--381.
\newline\urlprefix\url{https://doi.org/10.1214/12-EJS675}
\endbibitem

\bibitem[{Le{\'o}n-Novelo and Savitsky(2019)}]{leon2019fully}
Le{\'o}n-Novelo, L.~G. and Savitsky, T.~D. (2019).
\newblock \enquote{Fully Bayesian estimation under informative sampling.}
\newblock {\em Electronic Journal of Statistics\/}, 13(1): 1608--1645.
\endbibitem

\bibitem[{Lumley(2019)}]{lumley2019surveyR}
Lumley, T. (2019).
\newblock \enquote{survey: analysis of complex survey samples.}
\newblock R package version 3.35-1.
\endbibitem

\bibitem[{Makela et~al.(2018)Makela, Si, and Gelman}]{makela2018bayesian}
Makela, S., Si, Y., and Gelman, A. (2018).
\newblock \enquote{Bayesian inference under cluster sampling with probability
  proportional to size.}
\newblock {\em Statistics in medicine\/}, 37(26): 3849--3868.
\endbibitem

\bibitem[{Pfeffermann et~al.(1998{\natexlab{a}})Pfeffermann, Krieger, and
  Rinott}]{pfeffermann1998parametric}
Pfeffermann, D., Krieger, A.~M., and Rinott, Y. (1998{\natexlab{a}}).
\newblock \enquote{Parametric distributions of complex survey data under
  informative probability sampling.}
\newblock {\em Statistica Sinica\/}, 1087--1114.
\endbibitem

\bibitem[{Pfeffermann et~al.(1998{\natexlab{b}})Pfeffermann, Skinner, Holmes,
  Goldstein, and Rasbash}]{pfeffermann1998weighting}
Pfeffermann, D., Skinner, C.~J., Holmes, D.~J., Goldstein, H., and Rasbash, J.
  (1998{\natexlab{b}}).
\newblock \enquote{Weighting for unequal selection probabilities in multilevel
  models.}
\newblock {\em Journal of the Royal Statistical Society: series B (statistical
  methodology)\/}, 60(1): 23--40.
\endbibitem

\bibitem[{Rabe-Hesketh and Skrondal(2006)}]{rabe2006multilevel}
Rabe-Hesketh, S. and Skrondal, A. (2006).
\newblock \enquote{Multilevel modelling of complex survey data.}
\newblock {\em Journal of the Royal Statistical Society: Series A (Statistics
  in Society)\/}, 169(4): 805--827.
\endbibitem

\bibitem[{Williams and Savitsky(2020)}]{10.1214/18-BA1143}
Williams, M.~R. and Savitsky, T.~D. (2020).
\newblock \enquote{{Bayesian Estimation Under Informative Sampling with
  Unattenuated Dependence}.}
\newblock {\em Bayesian Analysis\/}, 15(1): 57 -- 77.
\newline\urlprefix\url{https://doi.org/10.1214/18-BA1143}
\endbibitem

\bibitem[{Williams and Savitsky(2021)}]{williams2021}
--- (2021).
\newblock \enquote{Uncertainty Estimation for Pseudo-Bayesian Inference Under
  Complex Sampling.}
\newblock {\em International Statistical Review\/}, 89(1): 72--107.
\newline\urlprefix\url{https://onlinelibrary.wiley.com/doi/abs/10.1111/insr.12376}
\endbibitem

\end{thebibliography}
\end{document}